\documentclass[a4,12pt]{article}
\usepackage{amsmath,amsthm,amssymb}
\usepackage[all]{xy}
\usepackage[dvipdfmx]{graphicx}

\usepackage[top=24truemm,bottom=24truemm,left=25truemm,right=25truemm]{geometry}

\numberwithin{equation}{section}

\theoremstyle{definition}
\newtheorem{df}{Definition}[section]
\newtheorem{prop}[df]{Proposition}
\newtheorem{ex}[df]{Example}
\newtheorem{thm}[df]{Theorem}
\newtheorem{rem}[df]{Remark}

\newtheorem{lem}[df]{Lemma}
\newtheorem{conj}[df]{Conjecture}
\newtheorem{fact}[df]{Fact}

\newcommand{\Proof}{\proof}

\newcommand{\bra}[1]{\left\langle #1 \right| }
\newcommand{\ket}[1]{\left| #1 \right\rangle }
\newcommand{\braket}[2]{\left\langle #1 | #2 \right\rangle }
\newcommand{\tbra}[1]{\langle #1 | }
\newcommand{\tket}[1]{| #1 \rangle }
\newcommand{\tbraket}[2]{\langle #1 | #2 \rangle}
\newcommand{\NPb}{{\rlap{\raise.4ex\hbox{$\scriptscriptstyle\bullet$}}
\lower.4ex\hbox{$\scriptscriptstyle\bullet$}}}
\newcommand{\NPc}{{\rlap{\raise.4ex\hbox{$\scriptscriptstyle\circ$}}
\lower.4ex\hbox{$\scriptscriptstyle\circ$}}}
\newcommand{\seteq}{\mathbin{:=}}
\newcommand{\rseteq}{\mathbin{=:}}
\newcommand{\twopii}{2 \pi \sqrt{-1} }

\newcommand{\chargQ}{\mathcal{Q}}

\newcommand{\vl}{\vec{\lambda}}
\newcommand{\vm}{\vec{\mu}}
\newcommand{\vn}{\vec{\nu}}
\newcommand{\lo}{\lambda^{(1)}}
\newcommand{\lt}{\lambda^{(2)}}
\newcommand{\lN}{\lambda^{(N)}}
\newcommand{\mo}{\mu^{(1)}}
\newcommand{\mt}{\mu^{(2)}}
\newcommand{\mN}{\mu^{(N)}}
\newcommand{\vu}{\vec{u}}
\newcommand{\vv}{\vec{v}}
\newcommand{\vw}{\vec{w}}
\newcommand{\Xo}{X^{(1)}}
\newcommand{\Xt}{X^{(2)}}

\newcommand{\tX}{\tilde{X}}
\newcommand{\tXo}{\tilde{X}^{(1)}}
\newcommand{\tXt}{\tilde{X}^{(2)}}
\newcommand{\tL}{\tilde{\Lambda}}
\newcommand{\tLo}{\tilde{\Lambda}^1}
\newcommand{\tLt}{\tilde{\Lambda}^2}
\newcommand{\tP}{\tilde{P}}
\newcommand{\tK}{\tilde{K}}
\newcommand{\tN}{\tilde{N}}
\newcommand{\tPhi}{\tilde{\Phi}}
\newcommand{\tT}{\tilde{T}}
\newcommand{\overstar}[1]{\mathop{\overset{*}{ #1 }}}
\newcommand{\overwstar}[1]{\mathop{\overset{**}{ #1 }}}

\newcommand{\be}{\begin{equation}}
\newcommand{\ee}{\end{equation}}
\newcommand{\ba}{\begin{eqnarray}}
\newcommand{\ea}{\end{eqnarray}}
\title{Crystallization of deformed Virasoro algebra, 
Ding-Iohara-Miki algebra and 5D AGT correspondence}
\author{Hidetoshi Awata \and Hiroki Fujino \and Yusuke Ohkubo}
\date{}

\begin{document}

\maketitle

\begin{abstract}
	In this paper, 
	we consider the $q \rightarrow 0$ limit of the deformed Virasoro algebra and 
	that of the level 1, 2 representation of Ding-Iohara-Miki algebra. 
	Moreover, 5D AGT correspondence at this limit is discussed. 
	This specialization corresponds to the limit from Macdonalds functions 
	to Hall-Littlewood functions. 
	Using the theory of Hall-Littlewood functions, 
	some problems are solved. 
	For example, 
	the simplest case of 5D AGT conjectures is proven at this limit, 
	and we obtain a formula for the 4-point correlation function of a certain operator. 
\end{abstract}

\section{Introduction}

Macdonald symmetric functions are orthogonal functions with many good properties 
and applications to various fields. 
Recently, 
they have played an important role in 5-dimensional AGT conjectures.

AGT conjectures \cite{alday2010liouville} are a sort of dualities 
between 2-dimensional conformal field theories and 4-dimensional gauge theories. 
Their $q$-deformations are dualities 
between the deformed Virasoro/W algebra or the Ding-Iohara-Miki algebra \cite{Ding-Iohara, Miki}
and 5-dimensional gauge theories  
\footnote{The correspondence between the elliptic Virasoro algebra and 6-dimensional theory is also proposed \cite{Nieri:2015dts, Iqbal:2015fvd}. }. 
In the simplest case, 
the norm of the Whittaker vector of the deformed Virasoro algebra coincides 
with the Nekrasov formula for the 5-dimensional pure gauge theory \cite{AwataYamada1, Yanangida2014norm}. 
The deformed Virasoro algebra is very closely related to Macdonald functions. 
The singular vectors of its highest weight representation 
correspond to Macdonald functions 
with rectangular Young diagrams \cite{shiraishi1996quantum, Awata:1995zk}. 
The Whittaker vector can be explicitly written in terms of Macdonald functions \cite{Yanangida2014whittaker}. 
Moreover, as one of the indication of AGT correspondence, 
there exists an good orthogonal basis called AFLT basis \cite{alba2011combinatorial, Fateev:2011hq}, 
by which conformal blocks can be expanded 
so that their each factor reproduce Nekrasov factor. 
The AFLT basis in the case of 5D AGT conjectures is in the representation space 
of the Ding-Iohara-Miki algebra 
and can be expressed by generalized Macdonald functions \cite{awata2011notes}. 
By using it, 
it is shown that $q$-Dotsenko-Fateev integrals are 
also decomposed into the form of Nekrasov formula \cite{Zenkevich, MorozovZenkevich}. 
The relationship between the Ding-Iohara-Miki algebra and 5D Nakrasov formula
is made clear in \cite{awata2012quantum}.

Macdonald functions, 
which contain two parameters $q$ and $t$,  
are a generalization of orthogonal functions 
called Jack functions and Hall-Littlewood functions.  
The degenerate limit to Jack functions, 
$q \rightarrow 1$ with $t=q^{\beta}$,  
agree with the limit from 5D AGT conjectures to 4D one. 
Then the deformed Virasoro algebra reduces to the ordinary Virasoro algebra, 
and generalized Macdonald functions are specialized 
to Morozov-Smirnov's generalized Jack functions \cite{Ohkubo}. 
In this limit, 
there is a scenario of the proof of 4D AGT conjectures as Hubbard-Stratanovich duality 
with help of these functions \cite{ morozov2013finalizing, SU(3)GnJack}, 
and a combinatorial formula of generalized Jack functions 
for the expansion in the basis of Schur functions is discovered in \cite{Smirnov}.

In this paper, 
we investigate behavior at the limit to Hall-Littlewood functions, 
$q \rightarrow 0$,  
of the deformed Virasoro algebra and 
algebras $\langle X^{(i)}_n \rangle$ generated by certain operators $X^{(i)}_n$ ($i=1,\ldots,N$), 
which is obtained by the level $N$ representation of the Ding-Iohara-Miki algebra. 
Also 5D AGT conjectures are studied at this limit. 
In this case, 
phenomena turn simple and some problems can be solved. 
In particular, by virtue of theories of Hall-Littlewood functions, 
we can solve the conjecture that 
the PBW type vectors of the algebra $\langle X^{(i)}_n \rangle$ 
form a basis over its representation space when $N=2$. 
By this fact, 
it is also proven that in the generic $q$ case the PBW type vector forms a basis. 
Furthermore, 
we can obtain and prove an explicit formula (Theorem \ref{thm:main theorem}) for the 4-point correlation function of a certain operator $\tPhi(z)$: 
\begin{equation*}
\bra{\vw} \tPhi(z_2) \tPhi(z_1) \ket{\vu}  
= \sum_{\lambda} \left( \frac{u_1u_2z_1}{w_1w_2z_2} \right)^{|\lambda|} 
\frac{ \prod_{k=1}^{\ell(\lambda)} \left(1- t^{k-1} \frac{w_1w_2}{v_1v_2} \right) }{t^{2n(\lambda)} b_{\lambda}(t^{-1}) }. 
\end{equation*}
Here for a partition $\lambda$, 
$n(\lambda) \seteq \sum_{i\geq 1} (i-1) \lambda_i$ 
and $b_{\lambda}(t)$ is defined in Appendix \ref{sec: Macdonald and HL}. 
The function $\bra{\vw} \tPhi(z_2) \tPhi(z_1) \ket{\vu} $ can be 
calculated by generalized Hall-Littlewood functions in the same way as \cite{awata2011notes}. 
However, 
this formula is obtained by inserting the identity with respect to PBW type vectors.

We call works at this limit "crystallization" after one of the quantum groups \cite{Kashiwara1990}
since the parameter $q$ also represents the temperature in the RSOS model \cite{LukyanovPugai:1996Multipoint} 
which has symmetry of the deformed Virasoro algebra, 
and the limit $q\rightarrow 0$ can be associated with the zero temperature. 
Although our studies are mathematically different 
from the notion of the original crystal base of quantum groups, 
the physical meaning and the motivation to simplify phenomena are same. 
To investigate their mathematical relationship is interesting further studies. 
On the other hand, the 5D Nekrasov formula express 
the instanton partition function of theories on $\mathbb{R}^4 \times S^1$. 
The limit $q \rightarrow 0$ correspond to the limit $R \rightarrow \infty$, 
where $R$ means the radius of $S^1$. 
Thus under the naive concept that the circle of radius infinity 
converge to a strait line $\mathbb{R}$, 
the Nekrasov function at this limit $\tilde{Z}^{\mathrm{inst}}_{\mathrm{pure}}$ (\ref{eq:tildeZ}) may express 
a partition function of some theory on $\mathbb{R}^5$.

Incidentally, 
Macdonald functions are reduced to Uglov functions \cite{Uglov:1997ia}
at the root of unity limit of parameters $q$ and $t$. 
AGT conjectures at this limit have been also studied. 
For example, 
the super Virasoro algebra are generated from the deformed Virasoro algebra 
at the limit $q, t \rightarrow -1$,  
and it correspond to theories on the ALE space $\mathbb{R}^4/\mathbb{Z}_2$ \cite{Itoyama:2013mca, Itoyama:2014qVir/W}. 
Moreover, 
certain conformal algebras $\mathcal{A}(r,k)$ are introduced in \cite{Belavin2013Instanton} 
and applied to AGT correspondence. 
The construction of these algebra is also investigated 
via the root of unity limit from the Ding-Iohara-Miki algebra in \cite{BBT2013Bases, Spodyneiko:2014qsa}. 
Integral formulas for the solutions of the KZ equation can also be constructed 
from the limit $q \rightarrow 1$, $t \rightarrow -1$ of the deformed Virasoro algebra 
\cite{Yoshioka:2015voz}. 
\begin{center}
\includegraphics[width=15cm]{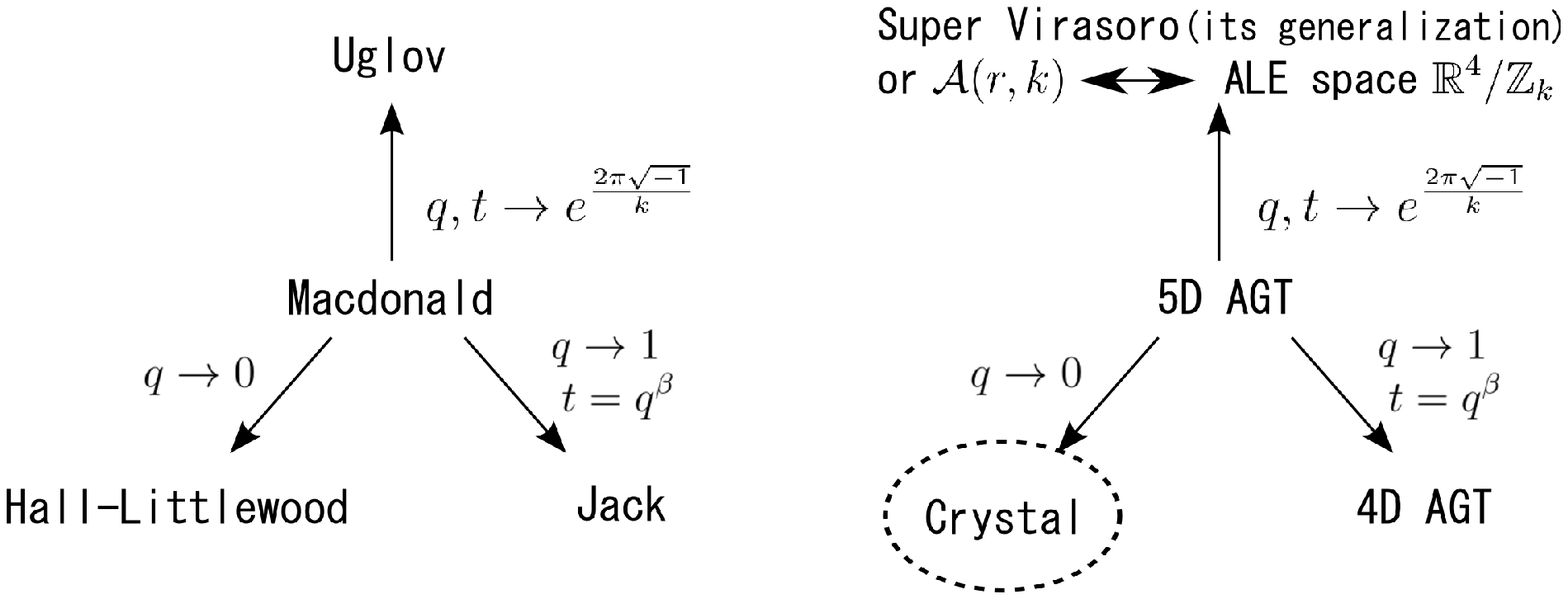}
\end{center}

This paper is organized as follows. 
We give a brief review on the deformed Virasoro algebra and the simplest 5D AGT correspondence 
in subsection \ref{sec:Review of Simplest 5D AGT}. 
In subsection \ref{sec:crystal of qVir}, 
we discuss its crystallization and 
prove the simplest 5D AGT correspondence at the $q \rightarrow 0$ limit.

In subsection \ref{sec:Reargument of DI alg and AGT}, 
we reargue the approach of AGT correspondence 
with the help of the level $N$ representation of the Ding-Iohara-Miki algebra and generalized Macdonald functions. 
In subsection \ref{sec:crystallization of N=1}, 
we give a conjecture at the $q\rightarrow 0$ limit in the case of $N=1$. 
In subsection \ref{sec:crystallization of N=2}, 
we consider a crystallization of the algebra $\langle X_n^{(i)} \rangle$ in the $N=2$ case. 
Next, 
we define generalized Hall-Littlewood functions 
and 
formulate conjectures of AGT correspondence at $q\rightarrow 0$. 
Moreover, 
we prove the formula for the correlation function obtained by PBW type vectors. 
In subsection \ref{sec:Another type of limit}, we describe different types of crystal limit
from the subsections \ref{sec:crystallization of N=1} and \ref{sec:crystallization of N=2}.

In the appendix, 
we describe the basic notion of Macdonald functions and Hall-Littlewood functions 
(subsection \ref{sec: Macdonald and HL}). 
Next, 
we explain several orderings which are required to state the existence theorem 
of generalized Macdonald functions (subsection \ref{sec:partial orderings}), 
and present some proofs and checks of conjectures in the text 
(subsection \ref{seq:Another proof of Lemma }-\ref{sec:check of N=1 conjecture}). 
Finally, we compare two formulas for the correlation function 
$\bra{\vw} \tPhi(z_2) \tPhi(z_1) \ket{\vu}$
obtained in subsection \ref{sec:crystallization of N=2} 
and provide a strange factorization formula arising from their comparison 
(subsection \ref{Comparison of two formula}). 


\section{Crystallization of the deformed Virasoro algebra and its Whittaker vector}

\subsection{Review of the simplest 5D AGT correspondence}\label{sec:Review of Simplest 5D AGT}

We start with recapitulating the resalt of the Whittaker vector of the deformed Virasoro algebra 
and the simplest 5-dimensional AGT correspondence.

\begin{df}
Let $q$ and $t$ be independent parameters and $p \seteq q/t$. 
The deformed Virasoro algebra is the associative algebra over $\mathbb{Q}(q,t)$ 
generated by $T_n$ ($n \in \mathbb{Z}$) with the commutation relation
\begin{equation}\label{comm.rel.of qVir}
[T_n, T_m] = -\sum_{l=1}^{\infty}f_l(T_{n-l}T_{m+l}-T_{m-l}T_{n+l})
             -\frac{(1-q)(1-t^{-1})}{1-p}(p^n-p^{-n}) \delta_{n+m,0},
\end{equation}
where the structure constant $f_l$ is defined by 
\begin{equation}
f(z) = \sum_{l=0}^{\infty}f_l \:  z^l
 \seteq \exp \left( \sum_{n=1}^{\infty}\frac{1}{n}\frac{(1-q^n)(1-t^{-n})}{1+p^n}z^n \right). 
\end{equation}
The relation (\ref{comm.rel.of qVir}) can be written 
in terms of the generating function $T(z)\seteq \sum_{n\in \mathbb{Z}}T_nz^{-n}$ as
\begin{equation}
f\left(\frac{w}{z}\right) T(z)T(w)-T(w)T(z)f\left(\frac{z}{w}\right) 
= -\frac{(1-q)(1-t^{-1})}{1-p}\left[ \delta\left( \frac{pw}{z} \right) - \delta\left( \frac{p^{-1}w}{z} \right) \right], 
\end{equation}
where $\delta(x)=\sum_{n\in \mathbb{Z}}x^n$.
\end{df}

Let $\ket{h}$ be the highest weight vector 
such that $T_0 \ket{h}=h \ket{h}$, $T_n\ket{h}=0$ ($n >0$), 
and 
$M_h$ be the highest weight module generated by $\ket{h}$. 
Similarly, 
$\bra{h}$ is the vector satisfying the condition that $\bra{h}T_0=h \bra{h}$, $\bra{h} T_{n}=0$ ($n< 0$). 
$M^{*}_h$ is the highest weight module generated by $\bra{h}$. 
The PBW theorem cannot be used because the deformed Virasoro algebra isn't Lie algebra.  
However, 
the PBW type vector $\tket{T_{-\lambda}} \seteq T_{-\lambda_1} T_{-\lambda_2} \cdots \ket{h}$ 
form a basis over $M_h$. 
Also, 
$\bra{T_{\lambda}} \seteq \bra{h} \cdots T_{\lambda_2} T_{\lambda_1}$ form a basis over $M^{*}_h$. 
Here $\lambda=(\lambda_1, \lambda_2, \ldots )$ is a partition or an Young diagram, i.e., 
a finite sequence of positive integers $\lambda_i$ such that $\lambda_{i} \geq \lambda_{i+1}$. 
Moreover, $\ell(\lambda)$ denote the length of $\lambda$, and $|\lambda| \seteq \sum_{i} \lambda_i$. 
The Whittaker vector $\ket{G}$ is defined as follows.

\begin{df}
Define the Whittaker vector
\footnote{The vector $\ket{G}$ is also called Gaiotto state. }
 $\ket{G}$ by the relations
\begin{equation}
T_1\ket{G}=\Lambda^2\ket{G}, \qquad T_n \ket{G}=0 \quad (n>1).
\end{equation}
Similarly, the dual Whittaker vector $\bra{G}\in M_h^*$ is defined by the condition that
\begin{equation}
\bra{G}T_{-1}=\Lambda^2\bra{G}, \qquad \bra{G} T_{n}=0 \quad (n<-1).
\end{equation}
\end{df}

This vector is in the form 
$\ket{G}=\sum_{\lambda}\Lambda^{2 |\lambda|} B^{\lambda, (1^n)} T_{-\lambda} \ket{h}$ 
and its norm is calculated as 
$\braket{G}{G} = \sum_{n=0}^{\infty} \Lambda^{4n} B^{(1^n),(1^n)}$, 
where for partitions $\lambda$ and $\mu$, $B^{\lambda, \mu}$ is the inverse matrix element of 
the Kac matrix $B_{\lambda, \mu} = \braket{T_{\lambda}}{T_{-\mu}}$.

It is useful to consider the free field representation of the deformed Virasoro algebra. 
By the Heisenberg algebra generated by $a_n$ ($n\in \mathbb{Z}$) and $\chargQ$ with the relations 
\begin{equation}
[a_n, a_m]=n\frac{1-q^{|n|}}{1-t^{|n|}} \delta_{n+m,0}, \qquad [a_n,\chargQ]=\delta_{n,0},
\end{equation}
the generating function $T(z)$ can be represented as 
\begin{align}\label{eq:free field rep.}
T(z) &= 
 \Lambda^+(z) + \Lambda^-(z), \\
\Lambda^{\pm}(z)
&:= \exp \left\{ \mp \sum_{n=1}^{\infty} \frac{1-t^{n}}{n(t^n+q^n)} (q/t)^{\mp\frac{n}{2}} a_{-n}z^n \right\} 
    \exp \left\{ \mp \sum_{n=1}^{\infty} \frac{1-t^n}{n} (q/ t)^{\pm \frac{n}{2}} a_{n}z^{-n} \right\} K^{\pm}. 
\end{align}
Here $K^{\pm} \seteq e^{\pm a_0}$. 
Let $\ket{0}$ be the highest weight vector such that $a_n\ket{k}=0$ ($n\geq 0$), 
and $\ket{k} \seteq k^{\chargQ}\ket{0}$. 
Then $K\ket{k}=k\ket{k}$. 
Furthermore, $\ket{k}$ can be regarded as the highest weight vector $\ket{h}$ 
of the deformed Virasoro algebra with highest weight $h=k+k^{-1}$. 
In \cite{Yanangida2014whittaker}, 
Yanagida proved an explicit formula for $\ket{G}$ in terms of Macdonald functions 
under the free field representation. 
The simplest 5-dimensional AGT conjecture is 
that the Norm $\braket{G}{G}$ corresponds to the 5-dimensional (K-theoretic) Nekrasov formula 
for pure gauge theory \cite{Awata:2008refined, NakajimaYoshioka1, NakajimaYoshioka2} : 
\begin{align}
Z^{\mathrm{inst}}_{\mathrm{pure}} &\seteq \sum_{\lambda, \mu} \frac{(\Lambda^4 t/q)^{|\lambda|+|\mu|}}{N_{\lambda \lambda}(1)N_{\lambda \mu}(Q)N_{\mu \mu}(1)N_{\mu \lambda}(Q^{-1})},  \\
N_{\lambda \mu}(Q) &\seteq \prod_{(i,j)\in \lambda} \left( 1- Q q^{A_{\lambda}(i,j)}t^{L_{\mu}(i,j)+1} \right)  \prod_{(i,j)\in \mu} \left( 1- Q q^{-A_{\mu}(i,j)-1} t^{-L_{\lambda}(i,j)} \right), 
\end{align}
where $A_{\lambda}(i,j) \seteq \lambda_i-j$ and $L_{\mu}(i,j)\seteq \lambda'_j-i $ are 
the arm length and the leg length of Young diagram, 
and $\lambda'$ is the conjugate of $\lambda$.

\begin{fact}\label{fact:simpleAGT}
 For $k=Q^{\frac{1}{2}}$,
\begin{equation}
\braket{G}{G}=Z^{\mathrm{inst}}_{\mathrm{pure}}.
\end{equation}
\end{fact}

This fact is conjectured in \cite{AwataYamada1} and proven in \cite{Yanangida2010five, Yanangida2014norm} for the generic $q$ case. 

\subsection{Crystallization of the deformed Virasoro algebra and AGT correspondence.}
\label{sec:crystal of qVir}

Next, we consider a crystallization of the results of the last subsection, that is 
behavior in the $q \rightarrow 0$ limit of the deformed Virasoro algebra 
and the simplest 5D AGT correspondence. 
In this limit, 
the scaled generators 
\begin{equation}
\tilde{T}_n \seteq (q/t)^{\frac{|n|}{2}}T_n 
\end{equation}
satisfy the commutation relation
\begin{align}
[\tilde{T}_n,\tilde{T}_m] =&
-(1-t^{-1})\sum_{\ell =1}^{n-m}{\tilde T }_{n-\ell}{\tilde T }_{m+\ell}
\quad (n > m > 0 \quad  \mbox{or}\quad  0>n>m),  
\\ 
[{\tilde T }_n,{\tilde T }_0 ] =&
 -(1-t^{-1})\sum_{\ell =1}^{n}{\tilde T }_{n-\ell}{\tilde T }_{\ell}
-(t-t^{-1})\sum_{\ell =1}^{\infty} t^{-\ell}
{\tilde T }_{-\ell}{\tilde T }_{n+\ell} \quad  (n > 0),  
\\ 
[{\tilde T }_0,{\tilde T }_m] =&
 -(1-t^{-1})\sum_{\ell =1}^{-m}{\tilde T }_{-\ell}{\tilde T }_{m+\ell}
-(t-t^{-1})\sum_{\ell =1}^{\infty} t^{-\ell}
{\tilde T }_{m-\ell}{\tilde T }_{\ell} \quad (0 > m),  \allowdisplaybreaks[4]
\\ 
[{\tilde T }_n,{\tilde T }_m] =&
-(1-t^{-1}){\tilde T }_{m}{\tilde T }_{n} 
-(t-t^{-1})\sum_{\ell =1}^{\infty} t^{-\ell}
{\tilde T }_{m-\ell}{\tilde T }_{n+\ell} \nonumber
\\ 
 & +(1-t^{-1})\delta_{n+m,0}
\quad (n> 0> m).
\end{align}

In \cite{awata1996Virasoro-tyme}, 
the above algebra is introduced 
and its free field representation is given. 
Let the bosons $b_n$ ($n \in \mathbb{Z}$) satisfy the relations 
$[b_n, b_m]=n \frac{1}{1-t^{|n|}}\delta_{n+m,0}$, 
$[b_n, \chargQ] = \delta_{n,0}$. 
These bosons can be regarded as the limit of the bosons $a_n$,  
i.e., $b_n= \lim_{q \rightarrow 0} a_n$, $\chargQ=\lim_{q \rightarrow 0} \chargQ$. 
Then $ {\tilde T }_n$ is represented as
\begin{equation}
{\tilde T }_n =\oint\frac{dz}{2 \pi \sqrt{-1} z}
\left( \theta[\,n\leq0\,] \tilde{\Lambda}^+(z) + 
           \theta[\,n\geq0\,] \tilde{\Lambda}^-(z)
\right) z^n,
\end{equation}
where
\begin{equation}
\tilde{\Lambda}^{\pm}(z)
 :=   \exp\left\{ \pm \sum_{n=1}^{\infty} \frac{1-t^{-n}}{n} b_{-n} z^{n} \right\}
       \exp\left\{ \mp \sum_{n=1}^{\infty} \frac{1-t^{n}}{n} b_{n} z^{-n} \right\} K^{\pm}
   = \lim_{q \rightarrow 0} \Lambda^{\pm}(p^{\pm 1/2} z) 
\end{equation}
and $\theta[P]$ is 1 or 0 if the proposition $P$ is true or false, respectively. 
By this free field representation, 
we can write the PBW type vectors in terms of Hall-Littlewood functions 
$Q_{\lambda}$ defined in Appendix \ref{sec: Macdonald and HL} : 
\begin{align}
\tT_{-\lambda} \ket{h} &= k^{\ell(\lambda)} Q_{\lambda}(b_{-n};t^{-1}) \ket{h}, \\
\bra{h} \tT_{\lambda} &= k^{-\ell(\lambda)} t^{|\lambda|}\bra{h} Q_{\lambda}(-b_n; t^{-1}).  
\end{align}
Here $Q_{\lambda}(b_{-n};t^{-1})$ is an abbreviation for $Q_{\lambda}(b_{-1}, b_{-2}, \ldots;t^{-1})$. 
This expression is shown by the theory of Jing's operator (Fact \ref{fact:Jing's operator}). 
Because of (\ref{eq:inner prod of HL poly}), they are diagonalized as 
\begin{equation}
\tilde{B}_{\lambda, \mu} 
\seteq \tbraket{\tT_{\lambda}}{ \tT_{\mu}} 
= \frac{1}{b_{\lambda}(t^{-1})} \delta_{\lambda, \mu}, 
\end{equation}
where $b_{\lambda}(t)$ is defined in Appendix \ref{sec: Macdonald and HL}. 
Since $\tilde{B}_{\lambda, \mu} $ is non-degenerate, 
there is no singular vector at $q \rightarrow 0$ limit. 
The disappearance of singular vectors can be understood by the fact 
that the highest weight which have singular vectors diverge at $q=0$. 
The Whittaker vector of this algebra is similarly defined.

\begin{df}
Define the Whittaker vector $\tket{\tilde{G}}$ by the relation
\begin{equation}
\tilde{T}_1\tilde{\ket{G}}=\tilde{\Lambda}^2 \tilde{\ket{G}}, \qquad \tilde{T}_n \tilde{\ket{G}}=0 \quad (n>1).
\end{equation}
Similarly, the dual Whittaker vector $\tilde{\bra{G}}\in M_h^*$ is defined by 
\begin{equation}
\tilde{\bra{G}}\tilde{T}_{-1}=\tilde{\Lambda}^2\tilde{\bra{G}}, \qquad \tilde{\bra{G}} \tilde{T}_{n}=0 \quad (n<-1).
\end{equation}
\end{df}

Then the crystallized Whittaker vector is in the simple form
\begin{equation}
\tket{\tilde{G}} 
= \sum_{\lambda} \tL^{2 |\lambda|}\, \tilde{B}^{\lambda, \mu} \, \tket{\tT_{\lambda}}
= \sum_{n=0}^{\infty} \tL^{2n}\, \frac{1}{ b_{(1^n)}(t^{-1}) } \, \tket{\tT _{-(1^n)}}, 
\end{equation}
and its inner product is 
\begin{equation}
\tbraket{\tilde{G}}{\tilde{G}} 
= \sum_{n=0}^{\infty} \tL^{4n} \tilde{B}^{(1^n), (1^n)}
= \sum_{n=0}^{\infty} \tL^{4n} \, \frac{1}{b_{(1^n)} (t^{-1})}. 
\end{equation}

On the other hand, we can take the limit of the Nekrasov formula as follows. 

\begin{prop}
The renormalization $\tL^2\seteq \Lambda^2 (q/t)^{\frac{1}{2}}$ 
controls divergence at the $q \rightarrow 0$ limit ($\Lambda \rightarrow \infty$, $\tL$ : fixed):   
\begin{equation}
\lim_{\substack{\Lambda^2=\tL^2(t/q)^{\frac{1}{2}} \\ q \rightarrow 0}}
Z^{\mathrm{inst}}_{\mathrm{pure}}
=\tilde{Z}^{\mathrm{inst}}_{\mathrm{pure}},
\end{equation}
\begin{equation}\label{eq:tildeZ}
\tilde{Z}^{\mathrm{inst}}_{\mathrm{pure}} 
\seteq \sum_{n,m\geq 0} 
{
\tilde \Lambda^{4(n+m)}
\over
\prod_{s=1}^n (1-t^{-s})(1-Q^{-1}t^{n-m-s})
\prod_{s=1}^m (1-t^{-s})(1-Q     t^{m-n-s})
}.
\end{equation}
\end{prop}

\proof
Removing parts which have singularity in the Nekrasov factor, we have
\begin{align}
N_{\lambda \mu}(Q) &=q^{-\sum_{(i,j)\in \mu}j} N'_{\lambda \mu}(Q),  \\
N'_{\lambda \mu}(Q) &:= 
  \prod_{(i,j)\in \lambda} \left( 1- Q q^{A_{\lambda}(i,j)}t^{L_{\mu}(i,j)+1} \right)  
  \prod_{(i,j)\in \mu} \left( q^{A_{\mu}(i,j)+1}- Q  t^{-L_{\lambda}(i,j)} \right).
\end{align}
Hence, 
\begin{equation}
Z^{\mathrm{inst}}_{\mathrm{pure}} 
= \sum_{\lambda, \mu} \frac{(\tilde{\Lambda}^4 t^2)^{|\lambda|+|\mu|} q^{E_{\lambda \mu}}}
{N'_{\lambda \lambda}(1)N'_{\lambda \mu}(Q)N'_{\mu \mu}(1)N'_{\mu \lambda}(Q^{-1})},
\end{equation}
\begin{equation}
E_{\lambda \mu} := 2\left( \sum_{(i,j) \in \lambda}j+ \sum_{(i,j) \in \mu} j -|\lambda|-|\mu| \right).
\end{equation}
If $\lambda \neq (1^n)$ or $\mu \neq (1^m)$ for any  integer $n$, $m$, 
then $q^{E_{\lambda \mu}} \rightarrow 0$ at $q \rightarrow 0$. 
Therefore, 
the sum with respect to partitions $\lambda$, $\mu$ can be rewritten 
as the sum with respect to integers $n$, $m$, 
i.e., 
\begin{align}
\tilde{Z}^{\mathrm{inst}}_{\mathrm{pure}} 
&= \sum_{n, m} \frac{(\tilde{\Lambda}^4 t^2)^{m+n} }
{\tilde{N}'_{nn}(1)\tilde{N}'_{nm}(Q)\tilde{N}'_{mm}(1)\tilde{N}'_{mn}(Q^{-1})}, \\
\tilde{N}'_{nm}(Q) &=(-1)^m Q^m t^{-nm+\frac{1}{2}m(m+1)} \prod_{s=1}^n \left( 1-Q t^{m-s+1} \right). 
\end{align}
After some simple algebra we get (\ref{eq:tildeZ}). 
\qed

Using these algebra and the Nekrasov function, 
we can get theorem which is an analog of Fact \ref{fact:simpleAGT} in the $q \rightarrow 0$ limit 
and prove it more easily than the generic case. 
\begin{thm}
\begin{equation}
\tbraket{\tilde{G}}{\tilde{G}}
= \tilde{Z}^{\mathrm{inst}}_{\mathrm{pure}}. 
\end{equation}
Note that the l.h.s. is independent of $k$. 
\end{thm}

\Proof
$\tilde{Z}^{\mathrm{inst}}_{\mathrm{pure}}= \tilde{Z}^{\mathrm{inst}}_{\mathrm{pure}}(Q)$ can be rewrite as 
\begin{equation}
\tilde{Z}^{\mathrm{inst}}_{\mathrm{pure}}(Q)
=\sum_{n,m\geq 0} 
{
\tilde \Lambda^{4(n+m)} Q^n t^{(n+1)m}
\over
\prod_{s=1}^n (1-t^{-s})(Q-t^{n-m-s})
\prod_{s=1}^m (1-t^{ s})(Q-t^{n-m+s})
},
\end{equation}
which has simple poles at
$Q=t^M$
with 
$-m\leq M\leq n$, $M\neq n-m$ and $M\in \mathbb{Z}$.
Then 
\begin{eqnarray}
{\rm Res}_{Q=t^M} \tilde{Z}^{\mathrm{inst}}_{\mathrm{pure}}(Q)
&=&
\sum_{n,m\geq 0} 
\tilde \Lambda^{4(n+m)} Z_{(n,m)}^{(M)}
,
\cr
Z_{(n,m)}^{(M)}
&:=&
{
t^{nM+(n+1)m} (t^M-t^{n-m})
\over
\prod_{s=1}^n (1-t^{-s})
\prod_{s=-m, (s\neq M)}^n (t^M-t^s)
\prod_{s=1}^m (1-t^{ s})
}
\cr
&=&
{
t^{(n-M)m} (t^{m+M}-t^{n})
\over
\prod_{s=1}^n (1-t^{-s})
\prod_{s=1}^{m+M} (1-t^{-s})
\prod_{s=1}^m (1-t^{ s})
\prod_{s=1}^{n-M} (1-t^{ s})
}.
\end{eqnarray}
Note that
\begin{equation}
Z_{(n,m)}^{(M)}
+
Z_{(m+M,n-M)}^{(M)} 
=0.
\end{equation}
Thus
\begin{equation}
Z_{\left( {N+M\over 2}+ r, {N-M\over 2}-r \right)}^{(M)}
=
{
t^{\left({N-M\over 2}\right)^2 + {N+M\over 2}-r^2 } (t^{-r}-t^{r})
\over
\prod_{s=1}^{{N+M\over 2}+r} (1-t^{-s})
\prod_{s=1}^{{N+M\over 2}-r} (1-t^{-s})
\prod_{s=1}^{{N-M\over 2}-r} (1-t^{ s})
\prod_{s=1}^{{N-M\over 2}+r} (1-t^{ s})
}
\end{equation}
is an odd function in $r$.
Therefore
\begin{equation}
{\rm Res}_{Q=t^M} \tilde{Z}^{\mathrm{inst}}_{\mathrm{pure}}(Q)
=
\sum_{N\geq 0} 
\tilde \Lambda^{4N} \sum_{r={|M|-N\over 2},(r\neq 0)}^{{N-|M|\over 2}}
Z_{\left( {N+M\over 2}+ r, {N-M\over 2}-r \right)}^{(M)}
=0.
\end{equation}
Residues at every singularities in $Q$ of 
$\tilde{Z}^{\mathrm{inst}}_{\mathrm{pure}}(Q)$ vanish, 
but $|\tilde{Z}^{\mathrm{inst}}_{\mathrm{pure}}(\infty)|<\infty$.
Hence $\tilde{Z}^{\mathrm{inst}}_{\mathrm{pure}}(Q)$ is independent of $Q$.
Therefore,
\begin{equation}
\tilde{Z}^{\mathrm{inst}}_{\mathrm{pure}}(Q)
=
\tilde{Z}^{\mathrm{inst}}_{\mathrm{pure}}(0)
=
\sum_{m\geq 0} 
{
\tilde \Lambda^{4m}
\over
\prod_{s=1}^m (1-t^{-s})
}.
\end{equation}

\qed

In this paper, 
we discuss the crystallization only of the deformed Virasoro algebra. 
It is expected that the limit can be taken for the general deformed $W_N$ algebra. 
However in the case of $W_3$, 
a pole like essential singularity appears 
and it isn't easy to control the singularity. 
To take the limit and apply the AGT conjecture 
for the deformed $W_N$ algebra \cite{Taki:2014fva}
is further studies. 

In the crystallized case, 
the screening current diverges, 
which is one of the reason why in this limit singular vectors disappear. 
Hence 
it may be difficult to apply the AGT correspondence studied by \cite{AwataYamada2}. 

\section{Crystallization of the representation of Ding-Iohara-Miki algebra}

\subsection{Reargument of Ding-Iohara-Miki algebra and AGT correspondence}
\label{sec:Reargument of DI alg and AGT}

We now turn to the Ding-Iohara-Miki algebra. 
First, we reargue the AFLT basis in the 5D AGT correspondence along \cite{awata2011notes}. 
In this section, 
we use $N$ kinds of bosons $a_n^{(i)}$ ($n \in \mathbb{Z}$, $i=1,2, \ldots , N$) 
and $\chargQ^{(i)}$ with the relations 
\begin{equation}
[a^{(i)}_n, a^{(j)}_m ]= n \frac{1-q^{|n|}}{1-t^{|n|}} \, \delta_{i,j} \, \delta_{n+m, 0}, 
\end{equation}
\begin{equation}
[a^{(i)}_n, \chargQ^{(j)}]=\delta_{i,j}\delta_{n,0}, 
\quad [\chargQ^{(i)}, \chargQ^{(j)}]=0 , \qquad (\forall i,j,n).
\end{equation}
Here the number $N \in \mathbb{N}$ corresponds to one of the $SU(N)$ gauge theory or the $W_N$ algebra. 
Let us define the vertex operators $\eta^{(i)}$ and $\varphi^{(i)}$ by 
\begin{align}
\eta^{(i)} (z)
&= \exp \left( \sum_{n=1}^{\infty} \frac{1-t^{-n}}{n}\, z^{n} a^{(i)}_{-n} \right) 
   \exp \left( -\sum_{n=1}^{\infty}\frac{(1-t^n)}{n} \, z^{-n} a^{(i)}_n \right), \\
\varphi^{(i)} (z)
&= \exp \left(  \sum_{n=1}^{\infty} \frac{1-t^{-n}}{n} (1-p^{-n})  z^{n} a^{(i)}_{-n} \right) ,
\end{align}
and introduce the following generators 
\begin{align}
 &X^{(1)}(z) \seteq \sum_{k=1}^N \Lambda^k(z) \rseteq \sum_{n} \Xo_n z^{-n}, \\
 &X^{(i)}(z) \seteq \Xo (p^{i-1} z) \cdots \Xo (p z) \Xo ( z) \rseteq \sum_{n} X^{(i)}_n z^{-n}, \nonumber
\end{align}
where
\begin{equation}
\Lambda^{i}(z) 
= \varphi^{(1)}(z p^{\frac{N-1}{2}}) \varphi^{(2)}(z p^{\frac{N-3}{2}}) \cdots \varphi^{(i-1)}(z p^{\frac{N-2i+3}{2}}) \eta^{(i)}(z p^{\frac{N-2i+1}{2}}) U_i, \quad 
U_i \seteq e^{a_0^{(i)}}. 
\end{equation}
The algebra generated by $X^{(i)}_n$ is 
obtained by the level $N$ representation of Ding-Iohara-Miki algebra \cite{FHHSY, FHSSY}. 
\footnote{
$X^{(i)}$ can be identified with one of \cite{awata2011notes} by the equation 
\begin{equation}\label{eq:relation of boson with AFHKSY}
p^{-\frac{N-k}{2}n} a_{n}^{(k)} = 1 \otimes \cdots \otimes a_{n} \otimes \dots \otimes 1. 
\end{equation}
In the $N=1,2$ case, 
singurality can be controled by this renormalization. 
Note that the commutation relations of $X^{(i)}(z)$ are not chainged.  
}

\begin{prop}
If $N=2$, 
the commutation relations of the generators are 
\begin{align}
&f^{(1)} \left( \frac{w}{z} \right) X^{(1)}(z) X^{(1)}(w) - X^{(1)}(w) X^{(1)}(z) f^{(1)} \left( \frac{z}{w} \right) \label{eq:rel. of generator X^1} \\ 
& \qquad \qquad \qquad = \frac{(1-q)(1-t^{-1})}{1-p} \left\{ \delta \left( \frac{w}{pz} \right) X^{(2)}(z ) -\delta \left( \frac{pw}{z} \right) X^{(2)}(w ) \right\} , \nonumber \\
&f^{(2)} \left( \frac{w}{z} \right) X^{(2)}(z) X^{(2)}(w) - X^{(2)}(w) X^{(2)}(z) f^{(2)} \left( \frac{z}{w} \right) =0,  \\
&f^{(1)} \left( \frac{pw}{z} \right) X^{(1)}(z) X^{(2)}(w) - X^{(2)}(w) X^{(1)}(z) f^{(1)} \left( \frac{z}{w} \right) =0,
\end{align}
where $\delta(x)=\sum_{n \in \mathbb{Z}} x^n$ is multiplicative delta function 
and  the structure constant $ f^{(i)}(z) =\sum_{n =0}^{\infty} f^{(i)}_l z^{l}$ is defined by 
\begin{equation}
 f^{(1)}(z) \seteq 
 \exp \left\{ \sum_{n >0} \frac{(1-q^n)(1-t^{-n})}{n}z^n \right\}, 
\end{equation}
\begin{equation}
 f^{(2)}(z) \seteq 
 \exp \left\{ \sum_{n >0} \frac{(1-q^n)(1-t^{-n})(1+p^{n})}{n}z^n \right\}. 
\end{equation}
These relations are equivalent to 
\begin{align}
& [X^{(1)}_n, X^{(1)}_m]= -\sum_{l =1}^{\infty} f^{(1)}_l (X^{(1)}_{n-l} X^{(1)}_{m+l} - X^{(1)}_{m-l} X^{(1)}_{n+l}) +\frac{(1-q)(1-t^{-1})}{1-p}(p^m-p^n) X^{(2)}_{n+m},  \\
& [X^{(2)}_n, X^{(2)}_m]= -\sum_{l =1}^{\infty} f^{(2)}_l (X^{(2)}_{n-l} X^{(2)}_{m+l} - X^{(2)}_{m-l} X^{(2)}_{n+l}), \\
& [X^{(1)}_n, X^{(2)}_m]= -\sum_{l =1}^{\infty} f^{(1)}_l (p^{l} X^{(1)}_{n-l} X^{(2)}_{m+l} -  X^{(2)}_{m-l} X^{(1)}_{n+l}) .
\end{align}
\end{prop}

In the formula (\ref{eq:rel. of generator X^1}) we use 
\begin{equation}
f^{(1)}(x) - f^{(1)}(1/px) = \frac{(1-q)(1-t^{-1})}{1-p} \left( \delta(x)- \delta(px) \right). 
\end{equation}

Let $\ket{0}$ and $\bra{0}$ be the vacuum state and its dual vector 
such that 
$a_n^{(i)} \ket{0}=0$ ($n \geq 0$, $\forall i$) and  
$\bra{0} a_n^{(i)} =0$ ($n \leq 0$, $\forall i$). 
For an N-tuple of parameters $\vu= (u_1, \ldots, u_N)$, 
define $\ket{\vec{u}} \seteq \prod_{i=1}^{N} u_i^{\chargQ^{(i)}} \ket{0}$ and  
$\bra{\vu} \seteq \bra{0} \prod_{i=1}^{N} u_i^{-\chargQ^{(i)}}$.  
Then $U_i\ket{\vec{u}}=u_i \ket{\vec{u}}$ and   
$\bra{\vec{u}} U_i=u_i \bra{\vec{u}}$. 
In this paper, let parameters $u_i$ be independent of $q$. 
$\mathcal{F}_{\vec{u}}$ is the highest weight module generated by $\ket{\vec{u}}$, 
and $\mathcal{F}_{\vec{u}}^*$ is the dual space generated by $\bra{\vec{u}}$. 
The PBW theorem can't be used because the algebra generated by $X^{(i)}_n$ isn't Lie algebra, 
but the following conjecture is given in \cite{awata2011notes}.

\begin{conj}
The PBW type vectors 
\begin{align}
 & X^{(1)}_{-\lambda^{(1)}_1} X^{(1)}_{-\lambda^{(1)}_2} \cdots  X^{(2)}_{-\lambda^{(2)}_1} X^{(2)}_{-\lambda^{(2)}_2}\cdots X^{(N)}_{-\lambda^{(N)}_1} X^{(N)}_{-\lambda^{(N)}_2} \cdots \ket{\vec{u}}\\
 & \left( \mathrm{resp. } \bra{\vec{u}} \cdots X^{(N)}_{\lambda^{(N)}_2} X^{(N)}_{\lambda^{(N)}_1} \cdots X^{(2)}_{\lambda^{(2)}_2} X^{(2)}_{\lambda^{(2)}_1} \cdots X^{(1)}_{\lambda^{(1)}_2} X^{(1)}_{\lambda^{(1)}_1} \right)
\end{align}
is a basis over $\mathcal{F}_{\vu}$ (resp. $\mathcal{F}_{\vec{u}}^*$), 
where $\vl=(\lo, \lt, \ldots, \lambda^{(N)})$ is an N-tuple of partitions. 
\end{conj}

In the $N=2$ case, 
we can solve this problem with respect to another PBW type vector 
by considering its crystallization. 
The $N=1$ case is also solved in the same way.

\begin{df}
For $\vl=(\lo, \lt, \ldots, \lambda^{(N)})$, set
\begin{align}
 &\ket{X_{\vec{\lambda}}} \seteq X^{(N)}_{-\lambda^{(N)}_1} X^{(N)}_{-\lambda^{(N)}_2} \cdots X^{(2)}_{-\lambda^{(2)}_1}X^{(2)}_{-\lambda^{(2)}_2}\cdots X^{(1)}_{-\lambda^{(1)}_1}X^{(1)}_{-\lambda^{(1)}_2}\cdots \ket{\vu}, \\
 & \bra{X_{\vec{\lambda}}} \seteq  \bra{\vec{u}} \cdots X^{(1)}_{\lambda^{(1)}_2} X^{(1)}_{\lambda^{(1)}_1} \cdots X^{(2)}_{\lambda^{(2)}_2} X^{(2)}_{\lambda^{(2)}_1} \cdots X^{(N)}_{\lambda^{(N)}_2} X^{(N)}_{\lambda^{(N)}_1}. 
\end{align}
\end{df}

\begin{prop}\label{prop:PBW vct form basis}
If $N=2$, then $\ket{X_{\vec{\lambda}}}$  (resp. $\bra{X_{\vec{\lambda}}}$ )
form a basis over $\mathcal{F}_{\vu}$ (resp. $\mathcal{F}_{\vec{u}}^*$). 
\end{prop}

The proof is given in the subsection \ref{sec:crystallization of N=2} 
(Remark \ref{rem:PBW in generic case form basis}).

Let us review the AFLT basis in $\mathcal{F}_{\vu}$, 
which is also called generalized Macdonald functions. 
In order to state its existence theorem, 
let us prepare the following ordering. 

\begin{df}\label{def:ordering1}
For N-tuple of partitions $\vl$ and $\vm$, 
\begin{align}
\vl \overstar{>} \vm \quad \overset{\mathrm{def}}{\Longleftrightarrow} \quad  
& |\vl| = |\vm|, \quad 
\sum_{i=k}^N |\lambda^{(i)}| > \sum_{i=k}^N |\mu^{(i)}| \quad (\forall k ) \quad \mathrm{and}   \\
&  (|\lo|,|\lt|,\ldots ,|\lN|) \neq (|\mo|,|\mt|,\ldots ,|\mN|).  \nonumber
\end{align}
Here $|\vl| \seteq |\lo|+\cdots + |\lambda^{(N)}|$. 
Note that the second condition can be replaced 
with $\sum_{i=1}^{k-1} |\lambda^{(i)}| < \sum_{i=1}^{k-1} |\mu^{(i)}|\quad  (\forall k )$. 
\end{df}

In the basis of products of Macdonald functions, 
we can state the existence theorem of generalized Macdonald functions.

\begin{prop}
For each N-tuple of partitions $\vl$, 
there exist an unique vector $\ket{P_{\vl}} \in \mathcal{F}_{\vu}$ 
such that 
\begin{align}
 &\ket{P_{\vec{\lambda}}} 
  = \prod_{i=1}^N P_{\lambda^{(i)}}(a^{(i)}_{-n};q,t) \ket{\vu} 
  + \sum_{\vm \mathop{\overset{*}{<}} \vl} c_{\vl, \vm} \prod_{i=1}^N P_{\mu^{(i)}}(a^{(i)}_{-n};q,t) \ket{\vu}, \\
 &X^{(1)}_0 \ket{P_{\vl}} = e_{\vl} \ket{P_{\vl}}, 
\end{align}
where $c_{\vl, \vm}=c_{\vl, \vm}(u_i,q,t)$ is a constant, 
$e_{\vl}=e_{\vl}(u_i,q,t)$ is an eigenvalue of $X^{(1)}_0$ 
and $P_{\lambda}(a_{-n}^{(i)};q,t)$ are Macdonald symmetric functions 
defined in Appendix \ref{sec: Macdonald and HL} 
with substituting the bosons $a^{(i)}_{-n}$ for power sum symmetric functions $p_n$. 
Similarly, there exist an unique vector $\bra{P_{\vl}} \in \mathcal{F}_{\vu}^*$ such that 
\begin{align}
 &\bra{P_{\vec{\lambda}}} 
  = \bra{\vu} \prod_{i=1}^N P_{\lambda^{(i)}}(a^{(i)}_{n};q,t) 
  + \sum_{\vec{\mu} \overstar{>} \vec{\lambda}} c_{\vl, \vm}^* \bra{\vu} \prod_{i=1}^N P_{\mu^{(i)}}(a^{(i)}_{n};q,t), \\
 & \bra{P_{\vec{\lambda}}} \Xo_0 = e_{\vec{\lambda}}^* \bra{P_{\vl}}.
\end{align}
\end{prop}

Although the ordering of Definition \ref{def:ordering1} 
is different from one of \cite{awata2011notes}, 
$\ket{P_{\vl}}$ coincides with one of \cite{awata2011notes}  
under the relation (\ref{eq:relation of boson with AFHKSY})
\footnote{If $N=2$, 
transformation $a^{(1)}_{-n}\ket{\vu} \mapsto p_n$, 
$a^{(2)}_{-n}\ket{\vu} \mapsto (q/t)^{n} \bar{p}_n$ 
and $\frac{u_1}{u_2}=Q$ 
maps $\ket{P_{\vl}}$ to $(q/t)^{|\lt|} M_{\vl}$ in \cite{Zenkevich}. 
}. 
The proof is similar to the one in the Appendix \ref{sec:partial orderings}, 
which follows from triangulation of $\Xo_0$. 
In Appendix \ref{sec:partial orderings}, 
a more elaborated ordering is introduced 
and relationship between these orderings is explained.

To use generalized Macdonald functions in the AGT correspondence, 
we need to consider its integral form. 
In this paper, 
we adopt following renormalization, 
which is slightly different from one of \cite{awata2011notes}.

\begin{df}\label{df:integral form of Gn Macdonald}
Define the vector $\ket{K_{\vec{\lambda}}}$ and $\bra{K_{\vec{\lambda}}}$, 
called the integral form, 
by the condition that 
\begin{align}
&\ket{K_{\vec{\lambda}}}= \sum_{\vm} \alpha_{\vl \vm} \ket{X_{\vm}} \propto \ket{P_{\vl}}, \quad 
\alpha_{\vl, (\emptyset,\ldots,\emptyset, (1^{|\vl|}))} =1, \\
&\bra{K_{\vec{\lambda}}}= \sum_{\vm} \beta_{\vl \vm} \bra{X_{\vm}} \propto \bra{P_{\vl}}, \quad 
\beta_{\vl, (\emptyset,\ldots, \emptyset, (1^{|\vl|}))} =1.
\end{align}
\end{df}

\begin{conj}
	The coefficients $\alpha_{\vl \vm}$ and $\beta_{\vl \vm}$ are 
	polynomials in $q^{\pm 1}$, $t^{\pm 1}$ and $u_i$ with integer coefficients.  
\end{conj}

\begin{ex}
	If $N=2$, 
	the transition matrix $\alpha_{\vl, \vm}$ is as follows:  
	\begin{equation*}
	\begin{array}{c||c c} 
	\vl \setminus \vm & (\emptyset, (1)) & ((1), \emptyset) \\ \hline \hline
	(\emptyset, (1))  & 1 &-\frac{q u_2}{t} \\
	((1), \emptyset)  & 1 &-\frac{q u_1}{t}
	\end{array},
	\end{equation*}
	\begin{equation*}
	\begin{array}{c||c c c } 
	\vl \setminus \vm & (\emptyset, (2)) & (\emptyset, (1^2)) &((1), (1)) \\ \hline \hline
	(\emptyset, (2)) & \frac{(q-1) u_2
		\left(t u_1 q^2-u_1 q^2+t u_2 q^2-u_2 q^2-u_2 q+tu_1\right)}{t^2} & 1  & -\frac{q (q+1) u_2}{t} \\
	(\emptyset, (1^2)) & \frac{q (t-1) u_2
		\left(-u_1 t^2+q u_2 t+q u_1-u_1+q u_2-u_2\right)}{t^3} & 1 & -\frac{q (t+1) u_2}{t^2}  \\
	((1), (1)) & \frac{(q-1) q (t-1)
		\left(u_1^2+u_2 u_1+u_2^2\right)}{t^2} & 1 & -\frac{q (u_1+u_2)}{t} \\
	((2), \emptyset) & \frac{(q-1) u_1
		\left(t u_1 q^2-u_1 q^2+t u_2 q^2-u_2 q^2-u_1 q+t
		u_2\right)}{t^2} & 1 & -\frac{q (q+1) u_1}{t} \\
	((1^2), \emptyset) & \frac{q (t-1) u_1
		\left(-u_2 t^2+q u_1 t+q u_1-u_1+q u_2-u_2\right)}{t^3} & 1  & -\frac{q (t+1) u_1}{t^2}
	\end{array},
	\end{equation*}
	\begin{equation*}
	\begin{array}{c|| c c} 
	\vl \setminus \vm  &((2), \emptyset) &((1^2), \emptyset) \\ \hline \hline
	(\emptyset, (2)) & -\frac{(q-1) q^2 u_2^2 (-q u_1+q t u_1+t u_1-q u_2)}{t^3} & \frac{q^3 u_2^2}{t^2}  \\
	(\emptyset, (1^2)) &-\frac{q^2 (t-1) u_2^2 (q u_1-t u_1-u_1+q u_2)}{t^4} & \frac{q^2 u_2^2}{t^3}  \\
	((1), (1)) & -\frac{(q-1) q^2 (t-1) u_1 u_2 (u_1+u_2)}{t^3} & \frac{q^2 u_1 u_2}{t^2} \\
	((2), \emptyset) & -\frac{(q-1) q^2 u_1^2 (-q u_1-q u_2+q t u_2+t u_2)}{t^3} &\frac{q^3 u_1^2}{t^2} \\
	((1^2), \emptyset) &-\frac{q^2 (t-1) u_1^2 (q u_1+q u_2-t u_2-u_2)}{t^4} & \frac{q^2 u_1^2}{t^3}
	\end{array}.
	\end{equation*}
\end{ex}

Then the norm of $\ket{K_{\vl}}$ reproduces the Nekrasov factor.

\begin{conj}
\begin{equation}
\tbraket{K_{\vl}}{K_{\vl}} 
\overset{?}{=} (-1)^N \, e_N(\vu)^{|\vl|} \, \prod_{i=1}^{N}t^{-N n(\lambda^{(i)})} q^{N n(\lambda^{(i)'})} u_i^{N |\lambda^{(i)}|}
 \prod_{i,j=1}^N N_{\lambda^{(i)}, \lambda^{(j)}} (qu_i/tu_j),
\end{equation}
where $e_N(\vu) = u_1 u_2 \cdots u_N$. 
\end{conj}

\begin{df}
Define the intertwining operator $\Phi(z) = \Phi^{\vv}_{\vu}(z): \mathcal{F}_{\vec{u}} \rightarrow \mathcal{F}_{\vec{v}}$ by 
\begin{equation}
(1-e_N(\vv) w/z) X^{(i)}(z) \Phi(w)= (1- p^{-i} e_N(\vv) w/z) \Phi(w) X^{(i)}(z)
\end{equation}
and $\bra{\vec{v}} \Phi(w) \ket{\vec{u}} = 1$. 
Then the relations for the Fourier components are 
\begin{equation}
(X^{(i)}_n- e_{N}(\vv) w X^{(i)}_{n-1})\Phi(w) 
= \Phi(w) (X^{(i)}_n-(t/q)^i e_{N}(\vv) w X^{(i)}_{n-1})
\end{equation}
for $i=1,2, \ldots , N$.
\end{df}

\begin{ex}
If $N=1$, it is known that $\Phi(z)$ has explicit form 
\begin{equation}
\Phi(z)= \exp \left\{ -\sum_{n=1}^{\infty} \frac{1}{n} \frac{v^n-(t/q)^n u^n}{1-q^n}a_{-n}z^n \right\}
\exp \left\{ \sum_{n=1}^{\infty} \frac{1}{n} \frac{v^{-n}- u^{-n}}{1-q^{-n}}a_{n}z^{-n} \right\} 
(v/u)^{\chargQ}. 
\end{equation} 
\end{ex}

\begin{conj}
The matrix elements of $\Phi(w)$ with respect to generalized Macdoald functions are
\begin{align}
\bra{K_{\vl}} \Phi^{\vv}_{\vu}(w) \ket{K_{\vm}}  
& \overset{?}{=} (-1)^{|\vl| +(N-1)|\vm|} (t/q)^{N(|\vl|-|\vm|)} e_N(\vu) ^{|\vl|} e_N(\vv)^{|\vl|-|\vm|} w^{|\vl|-|\vm|} \\ 
& \quad \times \prod_{i=1}^N u_i^{N |\mu^{(i)}|} q^{N\, n(\mu^{(i)'})} t^{-N\, n(\mu^{(i)})}
\times \prod_{i,j=1}^N N_{\lambda^{(i)},\mu^{(j)}}(qv_i/tu_j). \nonumber 
\end{align}
\end{conj}

Under these conjectures, 
we can obtain the formula for multi-point correlation functions of $\Phi(z)$, 
and the formula for 4-point functions agree with the 5D $U(N)$ Nekrsov formula with 4 matters. 
This M-theoretic derivation is also given by \cite{Tan2013}. 

\subsection{Crystallization of $N=1$ case}\label{sec:crystallization of N=1}

Next, 
we discuss a crystallization of the results of the last subsection. 
At first, let us demonstrate the $q\rightarrow 0$ limit in the $N=1$ case. 
In this subsection, let us use the same bosons $b_n$ and $\chargQ$ 
as subsection \ref{sec:crystal of qVir}. 
Since singularity in $\Phi(z)$ can be removed by normalization $\Phi(p z)$, 
define the vertex operator $\tPhi(z)$ by 
\begin{equation}
\tPhi(z)\seteq \lim_{q \rightarrow 0} \Phi(pz) 
= \exp \left\{ \sum_{n=1}^{\infty} \frac{u^n}{n} b_{-n}z^n \right\}
\exp \left\{ \sum_{n=1}^{\infty} \frac{1}{n} \frac{u^{-n}- v^{-n}}{t^{-n}}b_{n}z^{-n} \right\} 
(v/u)^{\chargQ}.  
\end{equation}
If $N=1$, $\tket{P_{\vl}}$ are ordinary Macdonald functions, 
and their integral form $\ket{K_{\lambda}}$ have, at $q=0$, the relation 
\begin{align}
 &\tket{\tK_{\lambda}} \seteq \lim_{q \rightarrow 0} \ket{K_{\lambda}} 
  = (-u/t)^{|\lambda|} t^{-n(\lambda)} Q_{\lambda}(b_{-n};t) \ket{u},  \\
 &\tbra{\tK_{\lambda}} \seteq \lim_{q \rightarrow 0} \bra{K_{\lambda}} 
  = (-u)^{|\lambda|} t^{-n(\lambda)} \bra{u}Q_{\lambda}(b_{n};t). 
\end{align}
Hence, the matrix elements $\tbra{\tK_{\lambda}}\tPhi(x) \tket{\tK_{\mu}}$ 
can be written in terms of integrals 
by virtue of the theory of Jing's operator $H(z)$ and $H^{\dagger}(z)$ 
defined in (\ref{eq:Jing's op}) and (\ref{eq:dual Jing's op}). 
Using the usual normal ordering product $\NPb \quad \NPb$ with respect to the bosons $b_n$ 
\footnote{
	Let $\mathcal{H}$ be the Heisenberg algebra generated by the bosons $b_n$ ($n\in \mathbb{Z}$), 
	$\chargQ$ and $1$.  
	$\mathcal{H}_c$ is the algebra obtained by making $\mathcal{H}$ commutative. 
	The normal ordering product $\NPb \quad \NPb$ is defined to be 
	the linear map from $\mathcal{H}_c$ to $\mathcal{H}$ 
	such that  
	for $\mathcal{P} \in \mathcal{H}_c$, 
	\begin{equation}
	\NPb \mathcal{P} b_n \NPb =
	  \left\{
	  \begin{array}{l}
	  \NPb \mathcal{P} \NPb \, b_n   , \quad n \geq 0,   \\
	  b_{n} \, \NPb \mathcal{P} \NPb , \quad n<0,
	  \end{array}
	  \right.
	\quad 
	\NPb \mathcal{P} \chargQ \NPb = \chargQ \NPb \mathcal{P} \NPb, 
	\end{equation}
	and $\NPb 1 \NPb =1$. 
	In the next subsection, 
	the same symbol $\NPb \quad \NPb$ denote 
	the normal ordering product with respect to the bosons $b^{(i)}_n$ 
	which is defined similarly. 
	}
, 
we have 
\begin{align}
& H^{\dagger}(w_{\ell(\lambda)}) \cdots H^{\dagger}(w_{1}) \tPhi(x) H(z_{1}) \cdots H(z_{\ell(\mu)}) \nonumber \\ 
& \qquad \qquad=\mathfrak{I}(w,x,z) \NPb H^{\dagger}(w_{\ell(\lambda)}) \cdots H^{\dagger}(w_{1}) \tPhi(x) H(z_{1}) \cdots H(z_{\ell(\mu)}) \NPb,\\
&\mathfrak{I}(w,x,z) \seteq \prod_{1 \leq j< i\leq \ell(\lambda)} \left( \frac{w_i-w_j}{w_i-tw_j} \right) 
    \prod_{1 \leq i< j\leq \ell(\mu)} \left( \frac{z_i-z_j}{z_i-tz_j} \right) 
    \prod_{\substack{1 \leq i \leq \ell(\lambda) \\ 1 \leq j \leq \ell(\mu)}} \left( \frac{w_i-tz_j}{w_i-z_j} \right)  \nonumber \\
 & \qquad \qquad \qquad \times \prod_{1 \leq i \leq \ell(\mu)} \left( \frac{x-(t/v)z_i}{x-(t/u)z_i} \right) 
    \prod_{1 \leq i \leq \ell(\lambda)} \left( \frac{w_i}{w_i-ux} \right). 
\end{align}
Thus
\begin{equation}
\tbra{\tK_{\lambda}}\tPhi(x) \tket{\tK_{\mu}} =
(-u)^{|\lambda|+|\mu|} t^{-n(\lambda)-n(\lambda)-|\mu|}
\oint\frac{dz}{2 \pi \sqrt{-1} z} \frac{dw}{2 \pi \sqrt{-1} w} \mathfrak{I}(w,x,z) z^{-\mu} w^{\lambda}, 
\end{equation}
where 
$\displaystyle \oint \frac{dz}{\twopii z} \frac{dw}{\twopii w} 
\seteq \oint \prod_{i=1}^{\ell(\mu)} \frac{dz_i}{2\pi \sqrt{-1}z_i} 
\prod_{i=1}^{\ell(\lambda)} \frac{dw_i}{2\pi \sqrt{-1}w_i}$, 
$z^{-\mu}\seteq z_1^{-\mu_1} \cdots z_{\ell(\mu)}^{-\mu_{\ell(\mu)}}$, 
$w^{\lambda}\seteq w_1^{\lambda_1} \cdots w_{\ell(\lambda)}^{\lambda_{\ell(\lambda)}}$, 
and  
the integration contour is $|w_{\ell(\lambda)}|> \cdots >|w_1|>|x|>|z_1|>\cdots |z_{\ell(\mu)}|$. 
This integral reproduce the $q \rightarrow 0$ limit of the Nekrasov factor.

\begin{df}
Set
\begin{align}
	\tilde{N}_{\lambda \mu}(Q) & \seteq \lim_{q \rightarrow 0} q^{n(\mu')} N_{\lambda \mu}((q/t)Q) \\
	&= (-Qt^{-1})^{|\check{\mu}|} \, t^{-\sum_{(i,j) \in \check{\mu}} L_{\lambda}(i,j)} 
	\prod_{(i,j)\in \mu \smallsetminus \check{\mu}} \left( 1- Q t^{-L_{\lambda}(i,j)-1} \right), 
	\nonumber
\end{align}
where $\check{\mu}$ is the set of boxes in $\mu$ whose arm-length $A_{\mu}(i,j)$ is not zero. 
For example, 
if $\mu=(5,3,3,1)$, $\check{\mu}=(4,2,2)$. 
This Nekrasov factor has 
the property $\tilde{N}_{\lambda \emptyset}(Q)=1$ for any $\lambda$.
\end{df}

Therefore, the conjecture in the crystallized case of $N=1$ is 
\begin{equation}\label{eq:N=1 conjecture}
\oint\frac{dz}{\twopii z} \frac{dw}{\twopii w} \mathfrak{I}(w,x,z) z^{-\mu} w^{\lambda}
\overset{?}{=} \tN_{\lambda, \mu} (v/u) x^{|\lambda|-|\mu|} u^{|\lambda|} (-v)^{|\mu|} t^{|\mu|+n(\lambda)}.
\end{equation}
The case of some particular partitions 
can be checked by calculating the contour integral 
(Appendix \ref{sec:check of N=1 conjecture}). 

\subsection{Crystallization of $N=2$ case }\label{sec:crystallization of N=2}

In this subsection, 
let us use the bosons $b^{(i)}_n$ ($n\in \mathbb{Z},\, i=1,2$) and $\chargQ^{(i)}$ with the relation
\begin{equation}
[b^{(i)}_n,b^{(j)}_m]=n\frac{1}{1-t^{|n|}} \delta_{i,j} \, \delta_{n+m, 0}, \quad 
[b_n^{(i)}, \chargQ^{(j)}] =0, 
\end{equation}
and regard $b^{(i)}_n=\lim_{q \rightarrow 0} a^{(i)}_n$, $\chargQ^{(i)} = \lim_{q \rightarrow 0} \chargQ^{(i)}$. 
Let us define the generator at $q \rightarrow 0$.

\begin{df}\label{def:crystal first generator}
Set
\begin{equation}
\tilde{X}^{(1)}_n \seteq \lim_{q \rightarrow 0} p^{\frac{|n|}{2}} X^{(1)}_n. 
\end{equation}
\end{df}

\begin{prop}\label{prop:free field rep of X^1}
Definition \ref{def:crystal first generator} is well-defined, 
i.e., $p^{\frac{|n|}{2}} X^{(1)}_n$ has no singularity at $q=0$, 
and its free field representation is 
\begin{equation}\label{eq:free field rep of tXo}
\tilde{X}^{(1)}_n=\oint \frac{dz}{2 \pi \sqrt{-1} z} 
\left\{ \theta [n\geq 0] \tilde{\Lambda}^{1}(z) + \theta [n\leq 0] \tilde{\Lambda}^{2}(z) \right\} z^n, 
\end{equation}
where $\theta$ is defined in Section \ref{sec:crystal of qVir} and
\begin{align}
&\tilde{\Lambda}^{1}(z) \seteq 
\exp \left\{ \sum_{n>0} \frac{1-t^{-n}}{n}z^n b^{(1)}_{-n} \right\}
\exp \left\{-\sum_{n>0} \frac{1-t^{n}}{n}z^{-n} b^{(1)}_{n} \right\}U_1, \\
 &\tilde{\Lambda}^{2}(z)
\seteq 
\exp \left\{-\sum_{n>0} \frac{1-t^{-n}}{n}z^n b^{(1)}_{-n} \right\}
\exp \left\{\sum_{n>0} \frac{1-t^{-n}}{n}z^n b^{(2)}_{-n} \right\}
\exp \left\{-\sum_{n>0} \frac{1-t^{n}}{n}z^{-n} b^{(2)}_{n} \right\}U_2. 
\end{align}
\end{prop}

\Proof
Define $\Lambda^1_n$ and $\Lambda^2_n$ by 
\begin{equation}
\Lambda^1(z) \rseteq \sum_{n \in \mathbb{Z}} \Lambda^1_n p^{-n/2}z^{-n} , \quad
\Lambda^2(z) \rseteq \sum_{n \in \mathbb{Z}} \Lambda^2_n p^{n/2}z^{-n}, 
\end{equation}
we can see $\Lambda^i_n$ is well-behaved in the limit $q \rightarrow 0$ by the form of $\Lambda^i(z)$. 
If $n>0$, 
\begin{equation}
\tilde{X}^{(1)}_n = \lim_{q \rightarrow 0}( \Lambda^1_n +\Lambda^2_n p^n )
=\lim_{q \rightarrow 0} \Lambda^1_n
= \oint \frac{dz}{2 \pi \sqrt{-1} z} \tilde{\Lambda}^1(z) z^n,  
\end{equation}
if $n<0$, 
\begin{equation}
\tilde{X}^{(1)}_n = \lim_{q \rightarrow 0}( \Lambda^1_n p^{-n} +\Lambda^2_n )
=\lim_{q \rightarrow 0} \Lambda^2_n
=\oint \frac{dz}{2 \pi \sqrt{-1} z} \tilde{\Lambda}^2(z) z^n,  
\end{equation}
and if $n=0$, 
\begin{equation}
\tilde{X}^{(1)}_n = \lim_{q \rightarrow 0}( \Lambda^1_0  +\Lambda^2_0 )
=\oint \frac{dz}{2 \pi \sqrt{-1} z} (\tilde{\Lambda}^1(z)+\tilde{\Lambda}^2(z) ).  
\end{equation}
Thus $\tilde{X}^{(1)}_n$ is well-defined 
and (\ref{eq:free field rep of tXo}) is the natural free field representation. 
\qed

For the second generator, the following rescale is suitable.

\begin{df}
Set
\begin{equation}
\tilde{X}^{(2)}_n \seteq \lim_{q \rightarrow 0} p^{\frac{n}{2}} X^{(2)}_n. 
\end{equation}
\end{df}

\begin{prop}
The free field representation of $\tilde{X}^{(2)}_n$ is obtained by 
\begin{equation}
\tilde{X}^{(2)}_n  = \oint \frac{dz}{2 \pi \sqrt{-1} z} \tXt(z) z^n,
\end{equation}
where
\begin{align}
& \tXt(z) = \NPb \tLo(z) \tLt(z) \NPb \\
 &= \exp \left\{ \sum_{n>0} \frac{1-t^{-n}}{n}z^n b^{(2)}_{-n} \right\}
\exp \left\{-\sum_{n>0} \frac{1-t^{n}}{n}z^{-n} b^{(1)}_{n} \right\}
\exp \left\{-\sum_{n>0} \frac{1-t^{n}}{n}z^{-n} b^{(2)}_{n} \right\}U_1 U_2. \nonumber
\end{align}
\end{prop}
This proposition is easily seen by checking $\lim_{q \rightarrow 0} X^{(2)}(p^{-1/2}z)$. 
We can calculate the commutation relation of these generators as follows.

\begin{prop}
The generators $\tilde{X}^{(1)}_n$ and $\tilde{X}^{(2)}_n$ satisfy the relations 
\begin{align}
[\tilde{X}^{(1)}_n, \tilde{X}^{(1)}_m] 
  &= -(1-t^{-1}) \sum_{l=1}^{n-m} \tilde{X}^{(1)}_{n-l} \tilde{X}^{(1)}_{m+l} 
  \qquad (n>m>0 \;\; \mathrm{or} \;\; 0>n>m), \\
[\tilde{X}^{(1)}_n, \tilde{X}^{(1)}_0] 
  &= -(1-t^{-1}) \sum_{l=1}^{n-1} \tilde{X}^{(1)}_{n-l} \tilde{X}^{(1)}_{l}  
     -(1-t^{-1}) \sum_{l=1}^{\infty} \tilde{X}^{(1)}_{-l} \tilde{X}^{(1)}_{n+l}  
     +(1-t^{-1}) \tilde{X}^{(2)}_{n}  
  \quad (n>0 ), \\
[\tilde{X}^{(1)}_n, \tilde{X}^{(1)}_m] 
  &= -(1-t^{-1})\sum_{l=0}^{\infty} \tilde{X}^{(1)}_{m-l} \tilde{X}^{(1)}_{n+l} + (1-t^{-1}) \tilde{X}^{(2)}_{n+m}  
   \quad (n>0>m ), \\
[\tilde{X}^{(1)}_0, \tilde{X}^{(1)}_m] 
  &= -(1-t^{-1}) \sum_{l=1}^{-m-1} \tilde{X}^{(1)}_{-l} \tilde{X}^{(1)}_{m+l}  
     -(1-t^{-1}) \sum_{l=1}^{\infty} \tilde{X}^{(1)}_{m-l} \tilde{X}^{(1)}_{l}  
     +(1-t^{-1}) \tilde{X}^{(2)}_{m}  
 \quad (0>m ),
\end{align}
\begin{align}
[\tilde{X}^{(1)}_n, \tilde{X}^{(2)}_m]  
  &= (1-t^{-1}) \sum_{l=1}^{\infty} \tilde{X}^{(2)}_{m-l} \tilde{X}^{(1)}_{n+l}
  \qquad (n>0,\; \forall m), \\
[\tilde{X}^{(1)}_0, \tilde{X}^{(2)}_m]  
  &= -(1-t^{-1}) \sum_{l=1}^{\infty} ( \tilde{X}^{(1)}_{-l} \tilde{X}^{(2)}_{m+l}- \tilde{X}^{(2)}_{m-l} \tilde{X}^{(1)}_{l}) 
  \quad ( \forall m),  \\
[\tilde{X}^{(1)}_n, \tilde{X}^{(2)}_m]  
  &= -(1-t^{-1}) \sum_{l=1}^{\infty} \tilde{X}^{(1)}_{n-l} \tilde{X}^{(2)}_{m+l}
  \qquad (n<0,\; \forall m),  
\end{align}
\begin{equation}
[\tilde{X}^{(2)}_n, \tilde{X}^{(2)}_m]
 =-(1-t^{-1}) \sum_{l=1}^{\infty} ( \tilde{X}^{(2)}_{n-l} \tilde{X}^{(2)}_{m+l}- \tilde{X}^{(2)}_{m-l} \tilde{X}^{(2)}_{n+l}) 
 \qquad (\forall n,m).
\end{equation}
\end{prop}

\Proof
These are obtained by the following relation of generating functions: 
\begin{align}
 \displaystyle g\left( \frac{w}{z} \right) \tLo(z) \tLo(w)- g \left( \frac{z}{w} \right) \tLo(w) \tLo(z)  &= 0,\\
 \displaystyle g\left( \frac{w}{z} \right) \tLt(z) \tLt(w)- g \left( \frac{z}{w} \right) \tLt(w) \tLt(z)  &= 0,\\
 \displaystyle \tLo(z) \tLt(w) + \left( g\left( \frac{z}{w} \right) -1-t^{-1} \right) \tLt(w) \tLo(z) 
&=   (1-t^{-1}) \delta \left( \frac{w}{z} \right) \NPb \tLo(z) \tLt(w) \NPb, \label{relation of tLo tLt}  
\\
 \displaystyle \tLo(z) \tXt(w) - g\left( \frac{z}{w} \right)\tXt(w) \tLo (z) &= 0, \\
 \displaystyle g\left( \frac{w}{z} \right) \tLt(z) \tXt(w) - \tXt(w) \tLt (z) &= 0, \\
 \displaystyle g\left( \frac{w}{z} \right) \tXt(z) \tXt(w) - g\left( \frac{z}{w} \right)\tXt(w) \tXt (z) &=0, 
\end{align}
where
\begin{equation}
g\left( x \right)=\exp \left\{ \sum_{n>0} \frac{1-t^{-1}}{n}x^n \right\} =1+ (1-t^{-1})\sum_{l=1}^{\infty} x^l, 
\end{equation}
and for (\ref{relation of tLo tLt}) we used the formula 
\begin{equation}
g(x)+g(x^{-1}) -1-t^{-1}= +(1-t^{-1}) \delta(x).
\end{equation}
\qed

The algebra generated by $\Xo_n$ and $\Xt_n$ is closely related to Hall-Littlewood functions. 
In particular,  
the PBW type vectors can be written as the product of two Hall-Littlewood functions.

\begin{df}
For a pair of partitions $\vec{\lambda}=(\lo, \lt)$, set
\begin{align}
 &\tket{\tX_{\vec{\lambda}}} \seteq \tX^{(2)}_{-\lambda^{(2)}_1} \tX^{(2)}_{-\lambda^{(2)}_2}\cdots \tXo_{-\lambda^{(1)}_1} \tXo_{-\lambda^{(1)}_2} \cdots \ket{\vec{u}},\\
 &\tbra{\tX_{\vec{\lambda}}} \seteq \bra{\vec{u}} \cdots \tX^{(1)}_{\lambda^{(1)}_2} \tX^{(1)}_{\lambda^{(1)}_1} \cdots \tX^{(2)}_{\lambda^{(2)}_2} \tX^{(2)}_{\lambda^{(2)}_1}.  
\end{align}
\end{df}

We have the expression of these vectors in terms of Hall-Littlewood polynomials.

\begin{prop}\label{prop:PBW vct Rep by Hall} 
\begin{align}
 &\tket{\tX_{\lambda, \mu}} = (u_1 u_2)^{\ell(\mu)} u_2^{\ell(\lambda)} Q_{\mu}(b^{(+)}_{-n};t^{-1}) Q_{\lambda}(b^{(-)}_{-n};t^{-1}) \ket{\vu}, \\
 &\tbra{\tX_{\lambda, \mu}} = u_1^{\ell(\lambda)} (u_1 u_2)^{\ell(\mu)} t^{|\lambda|+|\mu|} \bra{\vu} Q_{\lambda}(b^{(-)}_{n};t^{-1}) Q_{\mu}(b^{(+)}_{n};t^{-1}), 
\end{align}
where 
\begin{align}
 b^{(+)}_{n} & \seteq b^{(1)}_{n} + b^{(2)}_{n}, \quad  b^{(+)}_{-n} \seteq  b^{(2)}_{-n} \quad (n>0),  \\
 b^{(-)}_{n} & \seteq b^{(1)}_{n}, \quad  b^{(-)}_{-n} \seteq -b^{(1)}_{-n} + b^{(2)}_{-n} \quad (n>0). 
\end{align}
\end{prop}

The vectors $\tXo_{-\lambda^{(1)}_1} \tXo_{-\lambda^{(1)}_2} \cdots  \tX^{(2)}_{-\lambda^{(2)}_1} \tX^{(2)}_{-\lambda^{(2)}_2}\cdots \ket{\vec{u}}$ 
don't have such a good expression. 
This proposition is proven by the theory of Jing's operator. 
Then the vectors $\tket{\tX_{\vec{\lambda}}}$ are partially diagonalized as the following proposition. 
Furthermore, 
with the help of Hall-Littlewood functions, 
we can calculate the Shapovalov matrix $S_{\vl,\vm} \seteq \tbraket{\tX_{\vl}}{\tX_{\vm}}$ 
and its inverse $S^{\vl,\vm}$.

\begin{prop}\label{prop:Shapovalov in terms of HL poly}
We can represent $S_{\vl, \vm}$ by the inner product $\langle -,- \rangle_{0,t}$ 
of Hall-Littlewood functions defined in Appendix \ref{sec: Macdonald and HL} : 
\begin{align}
S_{\vl,\vm} =& (u_1 u_2)^{\ell (\lt) +\ell(\mt)} u_1^{\ell(\lo)} u_2^{\ell(\mo)} \label{eq:Shapovalov} \\
& \times \frac{1}{b_{\lo}(t^{-1})}  \left\langle Q_{\lt}(p_n;t^{-1}), Q_{\mt}(-p_n;t^{-1}) \right\rangle_{0,t^{-1}} 
\delta_{\lo, \mo}, \nonumber \\
S^{\vl,\vm} =&  (u_1 u_2)^{-\ell(\mt) -\ell(\lt)} u_1^{-\ell(\mo)} u_2^{-\ell(\lo)} \label{eq:inverse Shapovalov} \\
 &\times  \frac{b_{\mo}(t^{-1})}{b_{\lt}(t^{-1}) b_{\mt}(t^{-1})} 
 \left\langle Q_{\lt}(-p_n;t^{-1}), Q_{\mt}(p_n;t^{-1}) \right\rangle_{0,t^{-1}} \delta_{\lo, \mo}. \nonumber
\end{align}
\end{prop}

\Proof
(\ref{eq:Shapovalov}) follows from Proposition \ref{prop:PBW vct Rep by Hall}. 
(\ref{eq:inverse Shapovalov}) can be checked by the equation
\begin{equation}
\sum_{\mu} \frac{
	\left\langle Q_{\lambda}(p_n;t), Q_{\mu}(-p_n; t) \right\rangle_{0,t} 
	\left\langle Q_{\mu}(-p_n; t), Q_{\nu}(p_n;t) \right\rangle_{0,t}}
    {b_{\mu}(t)}
=b_{\lambda}(t) \delta_{\lambda, \nu}
\end{equation}
which is given by inserting the complete system with respect to $Q_{\mu}(-p_n;t)$ 
into the equation 
$\left\langle Q_{\lambda}(p_n;t), Q_{\nu}(p_n;t) \right\rangle_{0,t} = b_{\lambda}(t) \delta_{\lambda, \nu}$. 
\qed

Existence of the inverse matrix $S^{\vl \vm}$ leads 
linear independence of $\tket{\tX_{\vl}}$. 
Since there are the same number of linear independent vectors as the dimension of each level of $\mathcal{F}_{\vu}$, 
we can see that $\tket{\tX_{\vl}}$ forms a basis over $\mathcal{F}_{\vu}$.

\begin{prop}
If $t$ is not a root of unity and $u_1, u_2 \neq 0$, 
$\tket{\tX_{\vl}}$ (resp. $\tbra{\tX_{\vl}}$) 
is a basis over $\mathcal{F}_{\vu}$ (resp. $\mathcal{F}_{\vu}^*$). 
\end{prop}

\begin{rem}\label{rem:PBW in generic case form basis}
Since vectors which are linear dependent for the generic $q$ are not linear independent in any limit, 
we can show that the PBW type vectors $\ket{X_{\vl}}$ and $\bra{X_{\vl}}$ in generic $q$ case 
form a basis over $\mathcal{F}_{\vu}$ and $\mathcal{F}_{\vu}^*$, respectively
\footnote{
	In this paper, 
	the parameters $u_i$ is independent of $q$. 
	If not so,  
	$\ket{X_{\vl}}$ can be linear dependent. 
	To clarify when $\ket{X_{\vl}}$ form a basis in the case that $u_i \in \mathbb{Q}(q,t)$, 
	we need to investigate their Kac determinat. 
	}
. 
(Proposition \ref{prop:PBW vct form basis})
\end{rem}

Next, let us introduce generalized Hall-Littlewood functions 
which are specialization of generalized Macdonald functions 
and give conjectures of AGT correspondence at $q=0$.

\begin{df}\label{def:Gn Hall-Littlewood}
Define the vector $\tket{\tP_{\vl}}$ and $\tbra{\tP_{\vl}}$ as the $q \rightarrow 0$ limit of generalized Macdonald functions, i.e., 
\begin{equation}
\tket{\tP_{\vl}} \seteq \lim_{q \rightarrow 0} \tket{P_{\vl}}, \qquad 
\tbra{\tP_{\vl}} \seteq \lim_{q \rightarrow 0} \tbra{P_{\vl}}. 
\end{equation}
We call the vectors $\tket{\tP_{\vl}}$ 
generalized Hall-Littlewood functions. 
\end{df}

These are the eigenvectors of $\tilde{X}^{(1)}_0$: 
\begin{equation}
\tilde{X}^{(1)}_0 \tket{\tilde{P}_{\vec{\lambda}}} 
=\tilde{e}_{\vec{\lambda}} \tket{\tilde{P}_{\vec{\lambda}}}, \quad 
\tbra{\tilde{P}_{\vec{\lambda}}} \tilde{X}^{(1)}_0  
=\tilde{e}^*_{\vec{\lambda}} \tbra{\tilde{P}_{\vec{\lambda}}}
\end{equation}
Moreover the eigenvalues are
\begin{equation}
\tilde{e}_{\vl} =\tilde{e}^*_{\vl} 
=\sum_{k=1}^2 u_k \left( 1+(1-t)\sum_{i\geq 1}^{\ell(\lambda^{(k)})}t^{-i}  \right). 
\end{equation}
However there are too many degenerate eigenvalues 
to ensure the existence of generalized Hall-Littlewood functions. 
It is difficult to characterize $\tket{\tilde{P}_{\vl}}$ 
as the eigenfunction of only $\tXo_0$. 
For example, 
$\vec{\lambda}=((1),(2))$ and $\vec{\mu}=((2),(1))$ 
have the relation $\vec{\lambda} \overstar{>} \vec{\mu}$, but
$\tilde{e}_{\vec{\lambda}} = \tilde{e}_{\vec{\mu}}$. 
\footnote{
Definition \ref{def:Gn Hall-Littlewood} is given under the hypothesis 
that the vector $\tket{P_{\vl}}$ has no singulality in the limit $q \rightarrow 0$. 
If we can show the existence therem of both generalized Macdonald 
and generalized Hall-Littlewood functions 
by using the same partial ordering and the same basis, 
this hypothesis is guaranteed. 
}

\begin{ex} 
The transition matrix $\tilde{c}_{\vl, \vm}$ is as follows,  
where 
\begin{equation}
\tket{\tP_{\vl}}= \sum_{\vm} \tilde{c}_{\vl, \vm} 
\prod_{i=1}^2 P_{\mu^{(i)}}(a_{-n}^{(i)};t) \ket{\vu}.
\end{equation}
\begin{equation*}
\begin{array}{c||c c} 
\vl \setminus \vm & (\emptyset, (1)) & ((1), \emptyset) \\ \hline \hline
(\emptyset, (1)) & 1 & \frac{u_2}{u_1-u_2} \\ 
((1), \emptyset) & 0 & 1 \\ 
\end{array} ,
\end{equation*}
\begin{equation*}
\begin{array}{c||c c c c c} 
 \vl \setminus \vm & (\emptyset, (2)) & (\emptyset, (1^2)) &((1), (1)) &((2), \emptyset) &((1^2), \emptyset) \\ \hline \hline
(\emptyset, (2)) & 1 & 0 & 0 & \frac{u_2}{u_1-u_2} & 0 \\ 
(\emptyset, (1^2)) & 0 & 1 & \frac{u_2}{tu_1-u_2} & -\frac{u_2}{tu_1-u_2} & \frac{tu_2^2}{(u_1-u_2)(t u_1-u_2)} \\ 
((1), (1)) & 0 & 0 & 1 & -1 & -\frac{t(1+t)u_2}{-u_1+tu_2} \\ 
((2), \emptyset) & 0 & 0 & 0 & 1 & 0 \\ 
((1^2), \emptyset) & 0 & 0 & 0 & 0 & 1\\ 
\end{array}.
\end{equation*}
The transition matrix $\tilde{c}^*_{\vl, \vm}$ is as follows,  
where $\tbra{\tP_{\vl}} 
= \sum_{\vm} \tilde{c}^*_{\vl, \vm} 
\bra{\vu}\prod_{i=1}^2 P_{\mu^{(i)}}(a_{n}^{(i)};t)$. 
\begin{equation*}
\begin{array}{c||c c} 
\vl \setminus \vm  & ((1), \emptyset) &(\emptyset, (1)) \\ \hline \hline
((1),\emptyset) & 1 & -\frac{u_2}{u_1-u_2} \\
(\emptyset, (1)) & 0 & 1 \\ 
\end{array} ,
\end{equation*}
\begin{equation*}
\begin{array}{c||c c c c c} 
\vl \setminus \vm   &((2), \emptyset) &((1^2), \emptyset)&((1), (1)) &(\emptyset, (2)) & (\emptyset, (1^2)) \\ \hline \hline
((2), \emptyset) &1 & 0 & 1-t & -\frac{u_2}{u_1-u_2} & 0 \\
((1^2), \emptyset) & 0 & 1 & \frac{t u_2}{t u_2-u_1} & 0 & \frac{t u_2^2}{(u_1-u_2)(u_1-t u_2)} \\
((1), (1)) & 0 & 0 & 1 & 0 & -\frac{(t+1) u_2}{t u_1-u_2} \\
(\emptyset, (2)) & 0 & 0 & 0 & 1 & 0 \\ 
(\emptyset, (1^2)) & 0 & 0 & 0 & 0 & 1\\ 
\end{array}.
\end{equation*}

\end{ex}

Similarly to the case of generic $q$, 
We define the integral form of generalized Hall-Littlewood functions 
and give a conjecture of its norm.

\begin{df}
The integral form $\tket{\tilde{K}_{\vec{\lambda}}}$ and 
$\tbra{\tK_{\vl}}$ are defined by
\begin{align}
&\tket{\tK_{\vl}}= \sum_{\vm} \tilde{\alpha}_{\vl \vm} \tket{\tX_{\vm}} 
\propto \tket{\tilde{P}_{\vl}}, \quad
\tilde{\alpha}_{\vl, (\emptyset ,(1^{|\vl|}))}=1, \\
&\tbra{\tK_{\vl}} 
= \sum_{\vm} \tilde{\beta}_{\vl \vm} \tbra{\tX_{\vm}} 
\propto \tbra{\tilde{P}_{\vl}}, \quad
\tilde{\beta}_{\vl, (\emptyset ,(1^{|\vl|}))}=1. 
\end{align}
Note that the coefficients $\tilde{\alpha}_{\vl, ((1^{|\vl|}), \emptyset )}$ and 
$\tilde{\beta}_{\vl, ( (1^{|\vl|}), \emptyset )}$ can be zero at $q=0$. 
\end{df}

\begin{conj}\label{conj:inner prod of Gn HL}
\begin{equation}
\tbraket{\tilde{K}_{\vec{\lambda}}}{\tilde{K}_{\vec{\lambda}}} 
\overset{?}{=} (u_1 u_2)^{|\vec{\lambda}|} u_1^{2|\lambda^{(1)}|} u_2^{2|\lambda^{(2)}|} t^{-2\left( n(\lambda^{(1)})+n(\lambda^{(2)}) \right)}
\prod_{i,j=1}^{2} \tilde{N}_{\lambda^{(i)}, \lambda^{(j)}}(u_i/u_j).
\end{equation}
\end{conj}

Next, 
let us define the vertex operator at crystal limit. 

\begin{df}
The vertex operator $\tPhi(z) : \mathcal{F}_{\vu} \rightarrow \mathcal{F}_{\vv}$ is defined 
by the relations 
\begin{align}
&\tXo_n \tPhi(z)= \tPhi(z) \tXo_n -v_1\, v_2\, z\, \tPhi(z)\tXo_{n-1} \quad (n\leq 0), \\
&\tXo_n \tPhi(z)= \tPhi(z) \tXo_n \quad (n\geq 1), \\
&\tXt_n \tPhi(z)= \tPhi(z) \tXt_n -v_1\, v_2\, z\, \tPhi(z)\tXt_{n-1} \quad (\forall n), \label{eq:comm rel of tPhi and tXt}
\end{align}
\begin{equation}
\bra{\vv}\tPhi \ket{\vu}=1. 
\end{equation} 
\end{df}

This definition is obtained 
by the renormalization $\tPhi(z)=\Phi(p^{\frac{3}{2}}z)$.  
We give some simple properties of the vertex operator $\tPhi(z)$. 

\begin{prop}
\begin{equation}\label{eq:simple property1}
\tbra{\tX_{\vl}} \tPhi(z) \ket{\vu}=
\left\{
\begin{array}{ll}
(-v_1 v_2 u_1 u_2 z)^n  , &\quad \vl=(\emptyset, (1^n)) \quad \mbox{for some $n$}, \\
0 , &\quad \mbox{otherwise.}
\end{array}
\right.
\end{equation}
For any $n \geq 1$, 
\begin{equation}
\tbra{\vv} \tPhi(z) \tX^{(i)}_{-n} \ket{\vu}=
\left\{
\begin{array}{ll}
\left(  \frac{1}{v_1v_2z} \right)^n (v_1+v_2+u_1+u_2) , &\quad i=1, \\
\left(  \frac{1}{v_1v_2z} \right)^n  (u_1 u_2-v_1v_2), &\quad i=2.
\end{array}
\right.
\end{equation}
\end{prop}

\Proof
These follow from the commutation relations of $\tPhi(z)$.
\qed

\begin{conj}\label{conj:mat element of phi wrt Gn HL}
The matrix elements of $\tPhi(z)$ with respect to the integral form $\tket{\tK_{\vl}}$ are
\begin{align}
\tbra{\tK_{\vl}} \tPhi(z) \tket{\tK_{\vm}} 
&\overset{?}{=} (-1)^{|\vl|+|\vm|} (u_1 u_2 v_1 v_2 z)^{|\vl|-|\vm|} u_1^{2|\mo|} u_2^{2|\mt|} (u_1u_2)^{|\vm|} t^{-2(n(\mo)+n(\mt))}\\
&\quad \times \prod_{i,j=1}^2 \tN_{\lambda^{(i)},\mu^{(j)}} (v_i/u_j). \nonumber
\end{align}
\end{conj}

Under these conjectures \ref{conj:inner prod of Gn HL} and
\ref{conj:mat element of phi wrt Gn HL}, 
we obtain the formula for correlation functions of the vertex operator $\tPhi(z)$. 
For example, the function corresponding to 4-point conformal blocks is  
\begin{align}
\bra{\vw} \tPhi_{\vv}^{\vw}(z_2) \tPhi_{\vu}^{\vv}(z_1) \ket{\vu}  
&= \sum_{\vl} \frac{\tbra{\vw} \tPhi(z_2) \tket{\tK_{\vl}}   \tbra{\tK_{\vl}} \tPhi(z_1) \tket{\vu}}{\tbraket{\tK_{\vl}}{\tK_{\vl}}} \nonumber \\
&\overset{?}{=} \sum_{\vl} \left( \frac{u_1u_2z_1}{w_1w_2z_2} \right)^{|\vl|}
\prod_{i,j=1}^{2} \frac{\tN_{\emptyset , \lambda^{(j)}} (w_i/v_j)   \tN_{\lambda^{(i)}, \emptyset} (v_i/u_j) }{\tN_{\lambda^{(i)}, \lambda^{(j)}} (v_i/v_j)} \nonumber \\
&= \sum_{\vl} \left( \frac{u_1u_2z_1}{w_1w_2z_2} \right)^{|\vl|}
\prod_{i,j=1}^{2} \frac{\tN_{\emptyset , \lambda^{(j)}} (w_i/v_j) }{\tN_{\lambda^{(i)}, \lambda^{(j)}} (v_i/v_j)}. \label{eq:expansion by AFLT} 
\end{align}
(\ref{eq:expansion by AFLT}) is AGT conjecture at $q \rightarrow 0$ limit with help of the AFLT basis. 
However, 
in the crystallized case, 
we can prove another formula for this 4-point correlation function 
by using the PBW type basis. 
At first, let us show following two lemmas. 

\begin{lem}\label{lem:matrix el. of tPhi wrt PBW}
The matrix elements with respect to PBW type vector $\tket{\tX_{\emptyset,\lambda}}$ 
and $\bra{\vv}$ are
\begin{equation}
\bra{\vv} \tPhi(z) \tket{\tX_{\emptyset,\lambda}}
=(-1)^{\ell (\lambda)} \left( \frac{1}{v_1v_2z} \right)^{|\lambda|} t^{-n(\lambda)} 
\prod_{k=1}^{\ell(\lambda)} (t^{k-1}v_1v_2-u_1u_2). 
\end{equation}
\end{lem}

\Proof
For $i \geq 2$, by (\ref{eq:comm rel of tPhi and tXt}) 
and the relation $\tXt_{-n+1} \tXt_{-n} = t^{-1} \tXt_{-n} \tXt_{-n+1}$, 
\begin{align}
\bra{\vv} \tPhi(z) \left( \tXt_{-i} \right)^m 
&= \left( \frac{1}{v_1v_2z} \right) \bra{\vv} \tPhi(z) \tXt_{-i+1} \left( \tXt_{-i} \right)^{m-1}    \\
&= \left( \frac{1}{v_1v_2z} \right) t^{-m+1} \bra{\vv} \tPhi(z) \left( \tXt_{-i} \right)^{m-1} \tXt_{-i+1}  \nonumber \\
&= \left( \frac{1}{v_1v_2z} \right)^m t^{-\frac{1}{2}m(m-1)} \bra{\vv} \tPhi(z) \left( \tXt_{-i+1} \right)^{m}.   \nonumber 
\end{align}
Repeating this calculation, 
\begin{equation}
\bra{\vv} \tPhi(z) \left( \tXt_{-i} \right)^m 
= \left( \frac{1}{v_1v_2z} \right)^{mk} t^{-\frac{1}{2}m(m-1)k} 
\bra{\vv} \tPhi(z) \left( \tXt_{-i+k} \right)^{m}, \label{eq:preparation for i>1}
\end{equation}
where  $0 \leq k\leq i-1$. 
When $i=1$, by similar calculation 
\begin{align}
\bra{\vv} \tPhi(z) \left( \tXt_{-1} \right)^m \tket{\vu} 
&=\left( -\frac{1}{v_1v_2z} \right) \left( v_1v_2 - t^{-m+1}u_1u_2 \right) \bra{\vv} \tPhi(z) \left( \tXt_{-1} \right)^{m-1} \tket{\vu}   \nonumber \\
&= \left( -\frac{1}{v_1v_2z} \right)^m \prod_{k=1}^{m}(v_1v_2-t^{-k+1}u_1u_2).  \label{eq:preparation for i=1}
\end{align}
By using above two formulas (\ref{eq:preparation for i>1}) and (\ref{eq:preparation for i=1}), 
if we write $\lambda=(i_1^{m_1}, i_2^{m_2}, \ldots , i_l^{m_l}) \; \; (i_1> i_2>\cdots >i_l)$, 
\begin{align}
\bra{\vv} \tPhi(z) \tket{\tX_{\emptyset, \lambda}}  
&= \left( \frac{1}{v_1v_2z} \right)^{m_1(i_1-i_2)} t^{-\frac{1}{2}m_1(m_1-1)(i_1-i_2)} \nonumber \\
& \quad \times \bra{\vv} \tPhi(z) \left( \tXt_{-i_2} \right)^{m_1+m_2} \left( \tXt_{-i_3} \right)^{m_3} \cdots \left( \tXt_{-i_l} \right)^{m_l} \ket{\vu} \nonumber \\
&= \left( \frac{1}{v_1v_2z} \right)^{m_1(i_1-i_3)+m_2(i_2-i_3)} t^{-\frac{1}{2}m_1(m_1-1)(i_1-i_2)-\frac{1}{2}(m_1+m_2)(m_1+m_2-1)(i_2-i_3)} \nonumber \\
& \quad \times \bra{\vv} \tPhi(z) \left( \tXt_{-i_3} \right)^{m_1+m_2+m_3} \left( \tXt_{-i_4} \right)^{m_4} \cdots \left( \tXt_{-i_l} \right)^{m_l} \ket{\vu} \nonumber \\
&= (-1)^{\ell(\lambda)} \left( \frac{1}{v_1v_2z} \right)^{|\lambda|} t^{-n(\lambda)} 
\prod_{k=1}^{\ell(\lambda)} (t^{k-1}v_1v_2-u_1u_2).
\end{align}
\qed

We have explicit form of parts of the inverse Shapovalov matrix.

\begin{lem}\label{lem:exlicit form of inv. shapovalov} 
\begin{equation}
S^{(\emptyset, (1^{|\lambda|})), (\emptyset,\lambda)}
=S^{(\emptyset,\lambda), (\emptyset, (1^{|\lambda|}))} 
= \frac{(-1)^{|\lambda|} t^{-n(\lambda)} (u_1u_2)^{-|\lambda|-\ell (\lambda)} }{b_{\lambda}(t^{-1})}. 
\end{equation}
\end{lem}

\Proof
In this proof, we put $s=|\lambda|$. 
Hall-Littlewood function $Q_{(1^{s})}(p_n;t)$ 
is the elementary symmetric function $e_{s}(p_n)$ times $b_{(1^{s})}(t)$. 
Elementary symmetric functions have the generating function 
\begin{equation}\label{eq:gen fn of elementary sym fn}
\sum_{k=0}^{\infty} z^{k} e_k(p_n) = \exp \left\{ -\sum_{n >0} \frac{(-z)^n}{n} p_n \right\}. 
\end{equation}
Hence by the $r=0$ case of Fact \ref{fact:spetialization of HL poly}, 
\begin{align}
\left\langle e_{s}(-p_n), Q_{\lambda}(p_n;t) \right\rangle_{0,t} 
&= \left. (-1)^s \exp \left\{ \sum_{n>0} \frac{z^n}{1-t^n} \frac{\partial}{\partial p_n} \right\} Q_{\lambda}(p_n;t) \right|_{\mathrm{coefficient}\; \mathrm{of}\; z^s} \\
 &= \left. (-1)^s Q_{\lambda}(p_n;t) \right|_{p_n \mapsto \frac{1}{1-t^n}}  \nonumber \\
 &= (-1)^s t^{n(\lambda)}. \nonumber
\end{align}
Therefore, 
the lemma follows from Proposition \ref{prop:Shapovalov in terms of HL poly}. 
\qed

We give other proofs of this lemma in Appendix \ref{seq:Another proof of Lemma }, 
and the form of $S^{(\emptyset, (|\lambda|)), (\emptyset,\lambda)}$ can be seen by 
Appendix \ref{sec:explicit form of <Q,Q>}. 
By the property (\ref{eq:simple property1}), Proposition \ref{prop:Shapovalov in terms of HL poly} and  
Lemmas \ref{lem:matrix el. of tPhi wrt PBW} and \ref{lem:exlicit form of inv. shapovalov}, 
we can show the following main theorem.

\begin{thm}\label{thm:main theorem}
\begin{align}\label{eq:expansion formula by PBW}
\bra{\vw} \tPhi(z_2) \tPhi(z_1) \ket{\vu}  
&= \sum_{\vl} \tbra{\vw} \tPhi(z_2) \tket{\tX_{\vl}} S^{\vl \vm}  \tbra{\tX_{\vm}} \tPhi(z_1) \tket{\vu} \nonumber \\
&= \sum_{\lambda} \tbra{\vw} \tPhi(z_2) \tket{\tX_{\emptyset, \lambda}} S^{(\emptyset,\lambda), (\emptyset,(1^{|\lambda|}))}  \tbra{\tX_{\emptyset, (1^n)}} \tPhi(z_1) \tket{\vu} \nonumber \\
&= \sum_{\lambda} \left( \frac{u_1u_2z_1}{w_1w_2z_2} \right)^{|\lambda|} 
   \frac{ \prod_{k=1}^{\ell(\lambda)} \left(1- t^{k-1} \frac{w_1w_2}{v_1v_2} \right) }{t^{2n(\lambda)} b_{\lambda}(t^{-1}) }. 
\end{align}
\end{thm}

In this way, 
the explicit formula for the correlation function can be obtained, 
where we don't use any conjecture. 
The formulas (\ref{eq:expansion by AFLT}) and (\ref{eq:expansion formula by PBW}) 
are compared in Appendix \ref{Comparison of two formula}. 
We expect that these works will be generalized to $N \geq3$ case. 

\subsection{Other types of limit}
\label{sec:Another type of limit}
Finally, let us present other types of the crystal limit. 
In this paper, 
we investigated the crystal limit while the parameters $u_i$, $v_i$ and $w_i$ are fixed. 
However, 
it is also important to study the cases that these parameters depend on $q$. 
For example, 
let us consider the case that $u_i=p^{-M_i} u'_i$, $v_i= p^{-A_i}v_i'$, $w_i=p^{-M_{i+2}} w_i'$ 
($M_i, A_i \in \mathbb{R}$) 
and $u_i'$, $v_i'$, $w_i'$ are independent of $q$ or fixed in the limit $q \rightarrow 0$. 
Let $M_i+1 > A_i > M_{j+2}$ for all $i, j\in \{1,2 \}$ 
and $A_1 =A_2$. 
Then the Nekrasov formula for generic $q$ case ($N=2$) 
\begin{equation}
Z^{\mathrm{inst}}_{N_f=4} \seteq 
\sum_{\vl} \left( \frac{u_1 u_2 z_1}{w_1 w_2 z_2} \right)^{|\vl|}
\prod_{i,j=1}^2 
\frac{N_{\emptyset, \lambda^{(j)}}(qw_i/t v_j) N_{\lambda^{(i)}, \emptyset}(qv_i/t u_j)}
{N_{\lambda^{(i)}, \lambda^{(j)}}(qv_i/t v_j)}
\end{equation}
depend only on the partitions of the shape $ \vl= ((1^n),(1^m))$ 
at the limit $q \rightarrow 0$, 
where $\left( \frac{u_1 u_2 z_1}{w_1 w_2 z_2} \right) =\tilde{\Lambda}$ is fixed, 
and coincides with the partition function of the pure gauge theory (\ref{eq:tildeZ}):  
\begin{equation}
Z^{\mathrm{inst}}_{N_f=4} \underset{q \rightarrow 0}{\longrightarrow} \tilde{Z}^{\mathrm{inst}}_{\mathrm{pure}},  
\end{equation}
where $Q=v_1'/v_2'$. 
Hence, we are sure that the vector 
\begin{equation}
\Phi(z) \ket{\vu}
\end{equation}
corresponds to the Whittaker vector in the section \ref{sec:crystal of qVir} at this limit, 
though we were not able to properly explain it. 
In this way, by considering the various other values of $M_i$ and $A_i$, 
we can find special behavior of $Z^{\mathrm{inst}}_{N_f=4}$ and the conformal block $\bra{\vec{w}} \Phi(z_2) \Phi(z_1)\ket{\vu}$ and may prove the relation. 
These are our future studies. 

\section*{acknowledgments}
	We would like to thank Professors 
	M. Bershtein, O. Foda, H. Itoyama, H. Kanno, Y. Matsuo, A. Morozov, 
	N. Nekrasov, M. Okado, K. Ohta, T. Oota and J. Shiraishi for discussions. 
	The work of H.A. is supported in part by Grant-in-Aid for Science Research
	[\# 24540210] from the Japan Ministry of Education, Culture, Sports, Science and Technology.  
	The work of Y.O. is supported in part by Grant-in-Aid for JSPS Fellow 26-10187. 

\appendix
\section{Appendix}

\subsection{Macdonald functions and Hall-Littlewood functions}\label{sec: Macdonald and HL}

In this subsection, 
we briefly review some properties of Hall-Littlewood functions and Macdonald functions.

Let $\Lambda_{N} \seteq \mathbb{Q}(q,t)[x_1, \ldots, x_N]^{S_N}$ 
be the ring of symmetric polynomials of $N$ variables and
$\Lambda \seteq \lim_{\leftarrow} \Lambda_N$ be the ring of symmetric functions. 
The inner product $\langle-,- \rangle_{q,t}$ over $\Lambda$ is defined such that 
for power sum symmetric functions $p_{\lambda} = \prod_{k\geq 1} p_{\lambda_k}$ 
($p_n =\sum_{i\geq 1} x_i^n$), 
\begin{equation}
\langle p_{\lambda}, p_{\mu} \rangle_{q,t} 
= z_{\lambda} \prod_{k=1}^{\ell(\lambda)} \frac{1-q^{\lambda_k}}{1-t^{\lambda_k}} \delta_{\lambda, \mu}, \quad 
z_{\lambda} \seteq \prod_{i \geq 1} i^{m_i} m_i !,
\end{equation}
where $m_i=m_i (\lambda)$ is the number of entries in $\lambda$ equal $i$. 
For a partition $\lambda$, 
Macdonald functions $P_{\lambda} \in \Lambda$ are uniquely determined 
by the following two conditions:
\begin{align}
 &\lambda  \neq \mu \quad \Rightarrow \quad \langle P_{\lambda}, P_{\mu} \rangle_{q,t}  = 0;  \\
 &P_{\lambda} = m_{\lambda} + \sum_{\mu < \lambda} c_{\lambda \mu} m_{\mu}.  
\end{align}
Here $m_{\lambda}$ is a monomial symmetric function and 
$<$ is the ordinary dominance partial ordering, 
which is defined as follows: 
\begin{equation}
\lambda \geq \mu \quad \overset{\mathrm{def}}{\Longleftrightarrow}  \quad 
\sum_{i=1}^k \lambda_i \geq \sum_{i=1}^k \mu_i \quad (\forall k) \quad \mathrm{and} \quad |\lambda| = |\mu|.
\end{equation}
In this paper, 
we regard power sum symmetric functions $p_n$ ($n \in \mathbb{N}$) 
as the variables of Macdonald functions, 
i.e., $P_{\lambda}= P_{\lambda}(p_n; q,t)$. 
Here $P_{\lambda}(p_n; q,t)$ is an abbreviation for $P_{\lambda}(p_1, p_2, \ldots; q,t)$. 
In this paper, we often use the symbol $P_{\lambda}(a_{-n}; q,t)$, 
which is the polynomial of bosons $a_{-n}$ obtained by replacing $p_n$ in Macdonald functions with $a_{-n}$.

Next, 
let the Hall-Littlewood function $P_{\lambda}(p_n;t)$ be given 
by $P_{\lambda}(p_n;t) \seteq P_{\lambda}(p_n;0,t)$. 
If $x_{N+1} = x_{N+2} = \cdots =0$, then 
for a partition $\lambda$ of length $\leq N$, 
Hall-Littlewood polynomials $P_{\lambda}(p_n;t)$ with $p_n=\sum_{i=1}^N x_i^n$
is expressed by 
\begin{equation}
P_{\lambda}(p_n;t) = \frac{1}{v_{\lambda}(t)} \sum_{w \in S_n} w \left( x_1^{\lambda_1} \cdots x_N^{\lambda_N} \prod_{i<j}\frac{x_i-tx_j}{x_i-x_j} \right), 
\end{equation}
where $v_{\lambda}(t) = \prod_{i\geq 0} \prod_{k=1}^{m_i(\lambda)}\frac{1-t^k}{1-t}$. 
Note that $m_0=N-\ell(\lambda)$. 
The action of the symmetric group $S_N$ of degree $N$ is defined by 
$w(x_1^{\alpha_1} \cdots x_N^{\alpha_N})= x_{w(1)}^{\alpha_1} \cdots x_{w(N)}^{\alpha_N}$ for $w \in S_N$.

It is convenient to introduce functions $Q_{\lambda}(p_n;t)$, 
which are defined by scalar multiples of $P_{\lambda}$ as follows:  
\begin{equation}
Q_{\lambda}(p_n;t) \seteq b_{\lambda}(t) P_{\lambda}(p_n;t), 
\end{equation}
where $b_{\lambda}(t) \seteq \prod_{i\geq 1} \prod_{k=1}^{m_i} (1-t^k)$. 
They are diagonalized as 
\begin{equation}\label{eq:inner prod of HL poly}
\langle Q_{\lambda}, Q_{\mu} \rangle_{0,t} = b_{\lambda}(t) \delta_{\lambda, \mu}, 
\end{equation}
and they are expressed by Jing's operator 
\begin{align}
H(z) &\seteq \exp \left\{ \sum_{n\geq 1} \frac{1-t^n}{n} b_{-n} z^n \right\} 
       \exp \left\{ -\sum_{n\geq 1} \frac{1-t^n}{n} b_{n} z^{-n} \right\} 
\rseteq \sum_{n \in \mathbb{Z}} H_n z^{-n}, \label{eq:Jing's op}\\
H^{\dagger}(z) &\seteq \exp \left\{ -\sum_{n\geq 1} \frac{1-t^n}{n} b_{-n} z^n \right\} 
       \exp \left\{ \sum_{n\geq 1} \frac{1-t^n}{n} b_{n} z^{-n} \right\}
\rseteq \sum_{n \in \mathbb{Z}} H^{\dagger}_n z^{-n}, \label{eq:dual Jing's op}
\end{align}
where $b_n$ is the bosons realized by 
\begin{equation}
b_{-n}=p_n, \qquad b_{n}= \frac{n}{1-t^n} \frac{\partial}{\partial p_n} \quad (n>0).
\end{equation}

\begin{fact}[\cite{Jing}] \label{fact:Jing's operator}
Let $\ket{0}$ be the vector such that $b_n \ket{0}=0$ ($n>0$). 
Then for a partition $\lambda$, we have
\begin{eqnarray}
&H_{-\lambda_1}H_{-\lambda_2} \cdots \ket{0} &= Q_{\lambda}(b_{-n};t)\ket{0}, \\
&\bra{0} \cdots H^{\dagger}_{\lambda_2}H^{\dagger}_{\lambda_1}  &= \bra{0} Q_{\lambda}(b_{n};t). 
\end{eqnarray}
\end{fact}

Furthermore, the following specialization formula is known.

\begin{fact}[\cite{Macdonald}]\label{fact:spetialization of HL poly}
Let $r$ be indeterminate. Under the specialization 
\begin{equation}
p_n = \frac{1-r^n}{1-t^n}, 
\end{equation}
Hall-Littlewood functions $Q_{\lambda} \in \Lambda$ is specialized as 
\begin{equation}
Q_{\lambda}(\frac{1-r^n}{1-t^n}; t) = t^{n(\lambda)} \prod_{i=1}^{\ell(\lambda)} (1-t^{1-i} r). 
\end{equation}
\end{fact}

\subsection{Partial orderings}\label{sec:partial orderings}

The existence theorem of generalized Macdonald functions can be stated 
by the ordering $\overstar{>}$ in Definition \ref{def:ordering1}. 
In this appendix, 
we introduce a more elaborated ordering. 
This ordering point out more elements which is $0$ in the transition matrix $c_{\vl, \vm}$, 
where $\ket{P_{\vl}} = \sum_{\vm} c_{\vl \vm} \prod_iP_{\mu^{(i)}}(a_{-n}^{(i)}) \ket{\vu} $, 
and give more strict condition to existence theorem.

\begin{df}
For N-tuples of partitions $\vl$ and $\vm$, 
\begin{align}
\vl \overstar{\succ} \vm \quad \overset{\mathrm{def}}{\Leftrightarrow} \quad
& \vl \overstar{>} \vm \quad \mathrm{and} 
& \{ \nu \mid \nu \supset \lambda^{(\alpha)}, \nu \supset \mu^{(\alpha)},  |\nu|=|\lambda^{(\alpha)}|+\sum_{\beta=\alpha+1}^N (|\lambda^{(\beta)}|-|\mu^{(\beta)}|) \} \neq \nonumber \emptyset
\end{align}
for all $\alpha$. 
Here $\lambda \supset \mu$ denote that $\lambda_i \geq \mu_i$ for all $i$. 
\end{df}

\begin{ex}
If $N=3$ and the number of boxes is $3$, 
then 
\[\xymatrix@!C=26pt{
  & (\emptyset, \emptyset, (3))  \ar[rd] & & (\emptyset, \emptyset, (2,1)) \ar[ld]\ar[rd] & & (\emptyset, \emptyset, (1,1,1))\ar[ld] & \\
  & & (\emptyset, (1), (2))\ar[lld]\ar[d]\ar[rrrrd] & & (\emptyset, (1), (1,1))\ar[lllld]\ar[d]\ar[rrd] & & \\
(\emptyset, (2), (1))\ar[d]\ar[rrd]\ar[rrrrd] & & ((1), \emptyset, (2))\ar[d] & & ((1), \emptyset, (1,1))\ar[lld] & & (\emptyset, (1,1), (1))\ar[d]\ar[lld]\ar[lllld] \\
(\emptyset, (3), \emptyset)\ar[d] & & ((1), (1), (1))\ar[lld]\ar[d]\ar[rrd]\ar[rrrrd] & & (\emptyset, (2,1), \emptyset)\ar[lllld]\ar[rrd]  & & (\emptyset, (1,1,1), \emptyset)\ar[d] \\
((1), (2), \emptyset)\ar[rrd]\ar[rrrrd] & & ((2), \emptyset, (1))\ar[d]  & & ((1,1), \emptyset, (1))\ar[d] & & ((1), (1,1), \emptyset)\ar[lllld]\ar[lld] \\
  & & ((2), (1), \emptyset)\ar[ld]\ar[rd] & & ((1,1), (1), \emptyset)\ar[ld]\ar[rd] & & \\
  & ((3), \emptyset, \emptyset) & & ((2,1), \emptyset, \emptyset) & & ((1,1,1), \emptyset, \emptyset). &
}\]
Here $\vl \rightarrow \vm$ stands for $\vl \overstar{\succ} \vm$. 
\end{ex}

By using the following conjecture, 
we can state the existence theorem.

\begin{conj}\label{conj:action of eta_n}
Let $\displaystyle \eta_n^{(i)} \seteq \oint \frac{dz}{2 \pi \sqrt{-1}z} \eta^{(i)}(z) z^{n}$.  
If $\eta_n^{(i)}$ ($n \geq 1$) act on Macdonald functions 
$P_{\lambda}(a_{-n}^{(i)};q.t) \ket{u}$, 
then there only appear partitions $\mu$ contained in $\lambda$, i.e.,  
\begin{equation}
\eta_{n}^{(i)} P_{\lambda}(a^{(i)}_{-n};q,t) \ket{u} = \sum_{\mu \subset \lambda} c_{\lambda, \mu} P_{\mu}(a^{(i)}_{-n};q,t) \ket{\vm}. 
\end{equation}
\end{conj}

\begin{thm}\label{thm: another existence thm}
Under the Conjecture \ref{conj:action of eta_n}, 
for an N-tuple of partitions $\vl$, 
there exist an unique vector $\ket{P_{\vl}} \in \mathcal{F}_{\vu}$ such that 
\begin{align}
 &\ket{P_{\vec{\lambda}}} 
  = \prod_{i=1}^N P_{\lambda^{(i)}}(a^{(i)}_{-n};q,t) \ket{\vu} 
  + \sum_{\vec{\mu} \overstar{\prec} \vec{\lambda}} c_{\vl, \vm} \prod_{i=1}^N P_{\mu^{(i)}}(a^{(i)}_{-n};q,t) \ket{\vu}, \\
 &X^{(1)}_0 \ket{P_{\vec{\lambda}}} = e_{\vec{\lambda}} \ket{P_{\vec{\lambda}}}. 
\end{align}
\end{thm}

\Proof
At first, 
$\eta_n^{(i)}$ satisfies 
\begin{align}
& \eta_0^{(i)} P_{\lambda}(a^{(i)}_{-n}) \ket{\vu}= \epsilon_{\lambda} P_{\lambda}(a^{(i)}_{-n}) \ket{\vu}, \quad 
\eta_0^{(j)} P_{\lambda}(a^{(i)}_{-n}) \ket{\vu}=  P_{\lambda}(a^{(i)}_{-n}) \ket{\vu}, \\
& \eta_n^{(j)} P_{\lambda}(a^{(i)}_{-n}) \ket{\vu}  \qquad  i \neq j, \quad n \geq 1. \nonumber
\end{align}
If we act $\displaystyle \Lambda_0^{i} \seteq \oint \frac{dz}{2 \pi \sqrt{-1} z} \Lambda^i(z)$ 
on the product of the Macdonald functions, then 
\begin{align}
\Lambda^{i}_0 \prod_{j=1}^{N} P_{\lambda^{(j)}}(a_{-n}^{(j)}) \ket{\vu} 
=&\epsilon_{\lambda^{(i)}} \prod_{j=1}^{N} P_{\lambda^{(j)}}(a_{-n}^{(j)}) \ket{\vu} \\
 &+ \sum_{\mu \subset \lambda^{(i)}} c'_{\lambda^{(i)}, \mu}(a_{-n}^{(1)},\ldots , a_{-n}^{(i-1)}) P_{\mu}(a_{-n}^{(i)}) \prod_{j \neq i} P_{\lambda^{(j)}}(a_{-n}^{(j)}) \ket{\vu},  \nonumber
\end{align}
where $c'_{\lambda^{(i)}, \mu}(a_{-n}^{(1)},\ldots , a_{-n}^{(i-1)})$ is a polynomial 
of degree
 $|\lambda^{(i)}| - |\mu|$ of $a_{-n}^{(1)},\ldots , a_{-n}^{(i-1)}$. 
\footnote{
 	That is to say, 
 	for the operator $\mathcal{O}$ such that 
 	$[\mathcal{O}, a^{(j)}_{-n} ]= n a^{(j)}_{-n}$, 
 	the polynomial $c'_{\lambda^{(i)}, \mu}$ satisfy 
 	$[\mathcal{O}, c'_{\lambda^{(i)}, \mu}] = (|\lambda^{(i)}| - |\mu|) c'_{\lambda^{(i)}, \mu}$. 
 }
Hence 
\begin{equation}
X_{0}^{(1)} \prod_{j=1}^{N} P_{\lambda^{(j)}}(a_{-n}^{(j)}) \ket{\vu} 
= e_{\vl} \prod_{j=1}^{N} P_{\lambda^{(j)}}(a_{-n}^{(j)}) \ket{\vu} 
 + \sum_{\vm \overstar{\prec} \vl} c_{\vl, \vm} \prod_{j=1}^{N} P_{\mu^{(j)}}(a_{-n}^{(j)}) \ket{\vu}. 
\end{equation}
Therefore one can easily diagonalize it and we have this theorem. 
\qed

In the basis of monomial symmetric functions 
$\ket{m_{\vl}} \seteq \prod_{i=1}^N m_{\lambda^{(i)}}(a_{-n}^{(i)}) \ket{\vu}$, 
we have 
\begin{align}
\Xo_0 \ket{m_{\vl}} 
 &= \Xo_0 \sum_{\vl \geq \vm} d_{\vl \vm} \prod_{i=1}^N P_{\mu^{(i)}}(a_{-n}^{(i)}) \ket{\vu} \\
 &= \sum_{\vl \geq \vm} d_{\vl \vm} \sum_{\vm \overstar{\succeq} \vn} d'_{\vl \vm} \prod_{i=1}^N P_{\nu^{(i)}}(a_{-n}^{(i)}) \ket{\vu}  \nonumber \\
 &= \sum_{\vl \geq \vm} \sum_{\vm \overstar{\succeq} \vn} \sum_{\vn \geq \vec{\rho}} d_{\vl \vm} \,  d'_{\vm \vn} \, d''_{\vn \vec{\rho}} \,  \ket{m_{\vec{\rho}}},   \nonumber 
\end{align}
where 
\begin{equation}
\vec{\lambda} \geq \vec{\mu} \quad \overset{\mathrm{def}}{\Leftrightarrow} \quad 
(|\lambda^{(1)}|, \ldots , |\lambda^{(N)}|) = (|\mu^{(1)}|, \ldots , |\mu^{(N)}|) 
\quad \mathrm{and} \quad
\lambda^{(\alpha)} \geq \mu^{(\alpha)}
\end{equation}
($1 \leq \forall \alpha \leq N$).
Thus the partial ordering $\overwstar{\succeq}$ defined as follows also triangulates $\Xo_0$. 
\begin{equation}
\vl \overwstar{\succeq} \vec{\rho} \quad \overset{\mathrm{def}}{\Leftrightarrow} \quad 
\mbox{there exsist $\vm$ and $\vn$ such that 
$\vec{\lambda} \geq \vec{\mu} \overstar{\succeq} \vec{\nu} \geq \vec{\rho}$ }. 
\end{equation}
It can be shown that the partial ordering $\overwstar{\succeq}$ 
is equivalent to the ordering $\geq^{\mathrm{L}}$ introduced in \cite{awata2011notes}. 
Therefore Theorem \ref{thm: another existence thm} support the existence theorem in \cite[Proposition3.8]{awata2011notes}.



\subsection{Other proofs of Lemma \ref{lem:exlicit form of inv. shapovalov}}
\label{seq:Another proof of Lemma }

In this subsection, 
let us explain other proofs of Lemma \ref{lem:exlicit form of inv. shapovalov} 
by the method of contour integrals.

The generating function of elementary symmetric functions and Jing's operator 
makes the equation
\begin{align}
& (-1)^s \left\langle e_{s}(-p_n), Q_{\lambda}(p_n;t) \right\rangle_{0,t}  \\
&= \oint \frac{dz}{\twopii z}\frac{dw}{\twopii w} 
    \prod_{i=1}^{l}\left( \frac{1}{1-zw_i}\right) 
    \prod_{1\leq i <j \leq \ell(\lambda)} \left( \frac{w_i-w_j}{w_i-tw_j} \right) 
    z^{-s}w^{-\lambda} \nonumber \\
 &\rseteq F_{\lambda_1, \lambda_2,\ldots , \lambda_{l}}, \nonumber
\end{align}
where we put $s=|\lambda|$, $l=\ell(\lambda)$, 
and $|1/z|>|w_1|>\cdots > |w_l|$. 
It suffices to show that $F_{\lambda_1, \lambda_2,\ldots , \lambda_{l}} = t^{n(\lambda)}$, 
which is proven by a recursive relation of $F$ as follows. 
The contour integral $\oint \frac{dw_1}{\twopii w_1}$ surrounding origin is represented 
as that surrounding $\infty$. 
Since $\lambda_1>0$, the residue of $w_1$ at $w_1=\infty$ is $0$. 
Hence, 
the only residue at $w_1=\frac{1}{z}$ is left, 
and it is
\begin{equation}
F_{\lambda_1, \lambda_2,\ldots , \lambda_{l}} 
 = \oint \frac{dz}{\twopii z} z^{-\lambda_2 \cdots -\lambda_l} 
   \prod_{i=2}^l \frac{dw_i}{\twopii w_i}w^{-\lambda_i} \prod_{i=2}^l \left( \frac{1}{1-tw_iz} \right)
   \prod_{2 \leq i <j \leq l} \left( \frac{w_i-w_j}{w_i-tw_j} \right). 
\end{equation}
By change of variable $t z \mapsto z$, 
\begin{equation}
F_{\lambda_1, \lambda_2,\ldots , \lambda_{l}} 
= \prod_{i=2}^{l} t^{\lambda_i} \cdot F_{\lambda_2, \ldots , \lambda_l}
\end{equation}
Therefore 
\begin{equation}
F_{\lambda_1, \lambda_2,\ldots , \lambda_{l}} 
= \left( \prod_{i=2}^{l} t^{\lambda_i}\right) \left(\prod_{i=3}^{l} t^{\lambda_i}\right) \cdots  t^{\lambda_l} 
\oint\frac{dz}{\twopii z} 
= t^{n(\lambda)}. 
\end{equation}
Thus the Lemma \ref{lem:exlicit form of inv. shapovalov} is proven.

Although it is slightly hard, 
one can also prove this lemma by reversing the order of integration, 
i.e., first perform over a variable $w_l$ surrounding origin.  
Indeed $w_l$ has the pole only at $w_l=0$, 
and its residue satisfies 
\begin{equation}
F_{\lambda_1, \ldots, \lambda_l}
=\sum_{\substack{\alpha_0, \alpha_1, \ldots , \alpha_{l-1} \geq 0  \\ \alpha_0+\cdots \alpha_{l-1}=\lambda_l }} 
 \left( \prod_{i=1}^{l-1} \mathcal{A}_{\alpha_i} \right) F_{\lambda_1+\alpha_1, \ldots , \lambda_{l-1}+\alpha_{l-1}}, 
\end{equation}
where for $n > 0$,  $\mathcal{A}_n \seteq (t-1)t^{n-1}$ and 
$\mathcal{A}_0 \seteq 1$. 
By the assumption that $F_{\beta} = t^{n(\beta)}$  for $\beta =(\beta_1, \beta_2, \ldots )$ 
with $\ell(\beta)=l-1$, 
we inductively  get
\begin{equation}
F_{\lambda_1, \ldots, \lambda_l}
=t^{n((\lambda_1, \ldots, \lambda_{l-1}))} 
\sum_{0 \leq k \leq \lambda_l} 
\sum_{\substack{\alpha_1, \ldots , \alpha_{l-1} \geq 0  
      \\ \alpha_1 +\cdots + \alpha_{l-1}= k }} 
 \left( \prod_{i=1}^{l-1} \mathcal{A}_{\alpha_i} \right) t^{n(\alpha)}. 
\end{equation}
By virtue of the equation 
\begin{equation}
\sum_{\substack{\alpha_1, \ldots , \alpha_{l} \geq 0  
      \\ \alpha_1 +\cdots + \alpha_{l}= k }} 
 \left( \prod_{i=1}^{l} \mathcal{A}_{\alpha_i} \right)
 t^{n(\alpha)}
= 
\left\{
\begin{array}{ll}
t^{lk}-t^{l(k-1)} ,  &\quad k\geq 1, \\
1 , &\quad k=0, 
\end{array}
\right.
\end{equation}
which is also proven by induction with respect to $l$, 
it can be seen that $F_{\lambda_1, \ldots , \lambda_l} = t^{n(\lambda)}$.  

\subsection{Explicit form of $\left\langle Q_{(s)}(-p_n), Q_{\lambda}(p_n;t) \right\rangle_{0,t}$}
\label{sec:explicit form of <Q,Q>} 

The formula for $S^{\lambda, (1^s)}$ is given 
in the Lemma \ref{lem:exlicit form of inv. shapovalov} and the last subsection. 
We also have an explicit form of $S^{\lambda, (s)}$ and $S_{\lambda, (s)}$. 
By the Proposition \ref{prop:Shapovalov in terms of HL poly}, 
it suffices to give the explicit form of 
$\left\langle Q_{(s)}(-p_n), Q_{\lambda}(p_n;t) \right\rangle_{0,t}$.

\begin{prop}
\begin{equation}
\left\langle Q_{(s)}(-p_n;t), Q_{\lambda}(p_n;t) \right\rangle_{0,t}
= t^{|\lambda|+n(\lambda)} \prod_{k=1}^{\ell(\lambda)} (1-t^{-k}). 
\end{equation}
\end{prop}

\Proof
The proof is similar to the previous subsection. 
Set 
\begin{equation}
G^k_{\lambda_1, \lambda_2, \ldots , \lambda_l}
\seteq \oint \frac{dz}{\twopii z} \frac{dw}{\twopii w} \prod_{i=1}^l \left( \frac{z-w_i}{t^{-k}z-tw_i} \right) 
\prod_{1\leq i<j \leq \ell(\lambda)} \left( \frac{w_i-w_j}{w_i-tw_j} \right)z^{|\lambda|} w^{-\lambda}. 
\end{equation}
Then 
$\left\langle Q_{(s)}(-p_n), Q_{\lambda}(p_n;t) \right\rangle_{0,t} 
= G^0_{\lambda_1, \lambda_2, \ldots , \lambda_l}$ 
by Jing's operator. 
Integration of $w_l$ around $\infty$ makes recursive relation
\begin{equation}
G^k_{\lambda_1, \lambda_2, \ldots , \lambda_l}
= t^{\lambda_1(k+1)-\ell(\lambda)}(t^{k+1}-1) G^{k+1}_{\lambda_2, \ldots , \lambda_l}, 
\end{equation}
and leads this Proposition. 
\qed

For general partitions $\lambda$ and $\mu$, 
we can get the integral representation of 
$\left\langle Q_{\lambda}(-p_n), Q_{\mu}(p_n;t) \right\rangle_{0,t}$. 
However, 
it is very hard to give their explicit formula. 

\subsection{Check of (\ref{eq:N=1 conjecture})}\label{sec:check of N=1 conjecture}

The integral formula (\ref{eq:N=1 conjecture}) can be checked
by the similar way to subsections \ref{seq:Another proof of Lemma } and \ref{sec:explicit form of <Q,Q>}. 

Let us set 
\begin{align}
\mathfrak{F}_{\lambda_1, \ldots, \lambda_l} (u) 
& \seteq 
\oint \prod_{i=1}^l \frac{dw_i}{\twopii w_i} w_i^{\lambda_i} 
\prod_{i=1}^l \left( \frac{w_i}{w_i-ux} \right) \prod_{1 \leq j< i \leq l} \frac{w_i-w_j}{w_i-tw_j},   \\
\mathfrak{G}_{\mu_1, \ldots, \mu_m} (u) 
& \seteq 
\oint \prod_{i=1}^m\frac{dz_i}{\twopii z_i} z_i^{-\mu_i} 
\prod_{1 \leq i< j\leq m} \left( \frac{z_i-z_j}{z_i-tz_j} \right)
\prod_{1 \leq i \leq m} \left( \frac{x-(t/v)z_i}{x-(t/u)z_i} \right). \nonumber
\end{align}
Then 
$\mathfrak{F}_{\lambda_1, \ldots, \lambda_l} (u) 
= (-u)^{-|\lambda|} t^{n(\lambda)} \tbra{\tK_{\lambda}}\tPhi(z) \tket{\tK_{\emptyset}}$, 
$\mathfrak{G}_{\mu_1, \ldots, \mu_l} (u) 
= (-u)^{-|\mu|} t^{n(\mu)+|\mu|} \tbra{\tK_{\emptyset}}\tPhi(z) \tket{\tK_{\mu}}$. 
The integration of $w_1$ around $0$ give the relation
\begin{equation}
\mathfrak{F}_{\lambda_1, \ldots, \lambda_l} (u) 
= (ux)^{\lambda_1} \mathfrak{F}_{\lambda_2, \ldots, \lambda_l} (tu). 
\end{equation}
On the other hand, 
the integration of $z_1$ around $\infty$ makes 
\begin{equation}
\mathfrak{G}_{\mu_1, \ldots, \mu_m} (u) 
= \left( 1-(u/v) \right) (ux/v)^{\mu_1} \mathfrak{G}_{\mu_2 \ldots, \mu_m} (u/t). 
\end{equation}
Thus 
\begin{align}
\mathfrak{F}_{\lambda_1, \ldots, \lambda_l} (u)  
&= (ux)^{\lambda_1} (utx)^{\lambda_2} \cdots (ut^{l-1}x)^{\lambda_l},  \\
\mathfrak{G}_{\mu_1, \ldots, \mu_m} (u)  
&= \left(1-\frac{u}{v}\right) \left(1-\frac{u}{vt}\right) \cdots \left(1-\frac{u}{vt^{m-1}}\right)
\left( \frac{u}{xt} \right)^{-\mu_1} \left( \frac{u}{xt^2} \right)^{-\mu_2} \cdots \left( \frac{u}{xt^m} \right)^{-\mu_m}.    \nonumber
\end{align}
These agree with the r.h.s. of (\ref{eq:N=1 conjecture}). 

\subsection{Comparison of formulas  (\ref{eq:expansion by AFLT}) and (\ref{eq:expansion formula by PBW})}
\label{Comparison of two formula}

In this subsection, we compare two formulas (\ref{eq:expansion by AFLT}) 
and (\ref{eq:expansion formula by PBW}) 
which are obtained by the other basis.

Comparison of the coefficients of $\frac{z_1}{z_2}$ gives the equation 
\begin{equation}\label{eq:comparison of two formula}
\sum_{|\vl |=n} 
\prod_{i,j=1}^{2} 
\frac{\tN_{\emptyset , \lambda^{(j)}} (w_i/v_j) }{\tN_{\lambda^{(i)}, \lambda^{(j)}} (v_i/v_j)}
\overset{?}{=}
\sum_{|\lambda|=n}
\frac{ \prod_{k=1}^{\ell(\lambda)} \left(1- t^{k-1} \frac{w_1w_2}{v_1v_2} \right) }{t^{2n(\lambda)} b_{\lambda}(t^{-1}) }. 
\end{equation}
Note that the l.h.s. is the summation with respect to pairs of partitions $\vl=(\lo, \lt)$ 
and the r.h.s. is the summation with respect to single partitions $\lambda$. 
The r.h.s. depend only on the ratio $\frac{w_1 w_2}{v_1 v_2}$ 
though the l.h.s. doesn't look that way.

For a single partition $\lambda$, 
let us define $\left\langle \lambda \right\rangle$ to be 
the set of all pairs of partitions $(\lo, \lt)$ such that 
a permutation of the sequence $(\lo_1, \lo_2, \ldots, \lt_1, \lt_2, \ldots)$ 
coincides with $\lambda$. 
For example, if $\lambda=(2,1,1)$, 
\begin{equation}
\left\langle \lambda \right\rangle 
= \{ ((2,1,1),\emptyset), ((2,1),(1)), ((2),(1,1)), ((1,1),(2)),((1),(2,1)), (\emptyset,(2,1,1)) \}. 
\end{equation}
Then we obtain the strange factorization formula  
with respect to the partial summation of l.h.s. in (\ref{eq:comparison of two formula})
\begin{equation}\label{eq:strange facorization formula}
\sum_{\vl \in \langle \lambda \rangle} 
\prod_{i,j=1}^{2} 
\frac{\tN_{\emptyset , \lambda^{(j)}} (w_i/v_j) }{\tN_{\lambda^{(i)}, \lambda^{(j)}} (v_i/v_j)}
\overset{?}{=}
\frac{ \prod_{k=1}^{\ell(\lambda)} \left(1- t^{k-1} \frac{w_1w_2}{v_1v_2} \right) }{t^{2n(\lambda)} b_{\lambda}(t^{-1}) }
\left(\frac{w_1w_2}{v_1v_2} \right)^{|\lambda|-\ell(\lambda)} t^{2 n(\lambda)-I_{\lambda}}, 
\end{equation}
where 
$I_{\lambda} \seteq \sum_{s\in \check{\lambda}}( L_{\emptyset}(s) -L_{\lambda}(s) )
= \sum_{(i,j)\in \check{\lambda}} \lambda'_j$. 
If we prove that l.h.s. of (\ref{eq:strange facorization formula}) 
depend only on $\frac{w_1w_2}{v_1v_2}$, 
(\ref{eq:strange facorization formula}) is easily seen 
by checking the case of $w_2=v_2$. 
This equation almost reproduce each term of the r.h.s. in (\ref{eq:comparison of two formula}). 
Hence (\ref{eq:comparison of two formula}) may be proven by this equation. 
If (\ref{eq:comparison of two formula}) holds, 
the AGT conjecture at $q \rightarrow 0$ with the help of the AFLT basis 
(\ref{eq:expansion by AFLT}) is completely proven. 


\noindent
HA, HF and YO: Graduate School of Mathematics, Nagoya University, Nagoya 464-8602, Japan 

\emph{E-mail address}\\
HA: \texttt{awata@math.nagoya-u.ac.jp} \\
HF: \texttt{m12040w@math.nagoya-u.ac.jp} \\
YO: \texttt{m12010t@math.nagoya-u.ac.jp}

\end{document}